\documentclass[a4paper,fleqn,usenatbib]{mnras}
\usepackage{amsmath}
\usepackage{txfonts}
\usepackage[T1]{fontenc}
\usepackage{ae,aecompl}
\usepackage{graphicx}
\usepackage{float}
\usepackage{gensymb}
\usepackage{upgreek}
\usepackage{myscrextend} 
\usepackage{caption}
\newcommand{\kms}{km\,s$^{-1}$}
\renewcommand{\ion}[2]{#1$\,${\small \MakeUppercase{\romannumeral #2}}}
\newcommand{\thisSN}{SN~2015U}

\title[The Type Ibn \thisSN]{\thisSN: A Rapidly Evolving and Luminous Type Ibn Supernova}
\author[Shivvers et al.]{Isaac Shivvers,$^{1}$\thanks{E-mail: ishivvers@berkeley.edu}
WeiKang Zheng,$^{1}$
Jon Mauerhan,$^{1}$
Io K. W. Kleiser,$^{1,2}$
\newauthor
Schuyler D. Van Dyk,$^{3}$
Jeffrey M. Silverman,$^{4}$
Melissa L. Graham,$^{1}$
\newauthor
Patrick L. Kelly,$^{1}$
Alexei V. Filippenko,$^{1}$
Sahana Kumar$^{1}$
\\
$^{1}$Department of Astronomy, University of California, Berkeley, CA 94720-3411, USA \\
$^{2}$Department of Astronomy, California Institute of Technology, Pasadena, CA 91125, USA \\
$^{3}$IPAC/Caltech, Mail Code 100-22, Pasadena, CA 91125, USA \\
$^{4}$Department of Astronomy, University of Texas, Austin, TX 78712, USA
}
\date{Submitted to MNRAS}
\pubyear{2016}
\begin{document}
\label{firstpage}
\pagerange{\pageref{firstpage}--\pageref{lastpage}}
\maketitle

\begin{abstract}
Supernova (SN) 2015U (also known as PSN~J07285387+3349106) was discovered in NGC~2388 on 2015 Feb. 11.
A rapidly evolving and luminous event, it showed effectively hydrogen-free spectra 
dominated by relatively narrow helium P-Cygni spectral features and it was
 classified as a SN~Ibn. In this paper we present photometric, spectroscopic, and spectropolarimetric observations of \thisSN,
including a Keck/DEIMOS spectrum (resolution $\approx 5000$) which fully resolves the optical emission and absorption features.
We find that \thisSN\ is best understood via models of shock breakout from extended
and dense circumstellar material (CSM), likely created by a history of mass loss from the progenitor with an extreme outburst
within $\sim1$--2\,yr of core collapse (but we do not detect any outburst in our archival imaging of NGC~2388).
We argue that the high luminosity of \thisSN\ was powered not through $^{56}$Ni decay but via the deposition
of kinetic energy into the ejecta/CSM shock interface.
Though our analysis is hampered by strong host-galaxy dust obscuration (which likely exhibits multiple components),
our dataset makes \thisSN\ one of the best-studied Type Ibn supernovae and provides a bridge of understanding to other
rapidly fading transients, both luminous and relatively faint.
\end{abstract}



\section{Introduction}
 
Core-collapse supernovae (SNe) are luminous explosions that mark the end of a massive star's life.
These SNe are differentiated from their thermonuclear counterparts (Type Ia SNe) and 
are grouped into classes via spectral and photometric analyses \citep[e.g.,][]{1997ARA&A..35..309F}.  The major 
core-collapse subclasses include SNe~IIP and IIL, events which show strong hydrogen throughout their evolution
and respectively do or do not exhibit a hydrogen-recombination plateau in their light curves
\citep[though the distinction between SNe IIP and IIL may be less clear than previously thought; e.g.,][]{2012ApJ...756L..30A,2014ApJ...786...67A,2015ApJ...799..208S}.
SNe Ib and Ic (often called stripped-envelope SNe) are core-collapse events that show no hydrogen in their spectra and 
(for SNe~Ic and broad-lined SNe~Ic) no helium; they (along with the intermediate SNe~IIb, which exhibit very little hydrogen) are commonly
understood to arise from progenitor stars that have lost all or most of their hydrogen (and perhaps helium) envelopes prior to
core collapse, though the detailed connections between progenitors and SN observables remain somewhat uncertain
\citep[e.g.,][]{2001AJ....121.1648M,2011ApJ...741...97D,2014ApJS..213...19B,2014AJ....147...99M,2015arXiv151008049L,2015MNRAS.453.2189D}.

A subset of SNe reveal signatures of a dense shroud of circumstellar material (CSM) surrounding their progenitors at the time of explosion.
For hydrogen-rich events, relatively narrow lines (full width at half-maximum intensity [FWHM] $\lesssim 1000$\,\kms)
from this interaction with the CSM are often detected; these objects have been dubbed SNe~IIn 
\citep[e.g.,][]{1990MNRAS.244..269S,1991ESOC...37..343F}.  
This interaction can provide a significant luminosity boost \citep[as has been observed for superluminous SNe~IIn; e.g.,][]{2012Sci...337..927G}.
CSM interaction occurs in some hydrogen-poor SNe as well.
SNe~Ia-CSM are thermonuclear events that exhibit relatively narrow features (usually H$\alpha$ emission)
caused by interaction with hydrogen-rich CSM \citep[though the underlying SN ejecta are hydrogen-poor; e.g.,][]{2013ApJ...772..125S,2013ApJS..207....3S}.
A small number of stripped-envelope SNe have been found to exhibit the spectral signatures of CSM interaction. SNe~Ibn
\citep[e.g.,][]{2007ApJ...657L.105F,2008MNRAS.389..113P} show relatively narrow helium lines in their spectra
but no hydrogen, and they can be quite heterogeneous in their photometric evolution \citep[e.g.,][]{2016MNRAS.456..853P},
while the remarkable SN~2014C appeared to be a normal SN~Ib at peak brightness but then began interacting with hydrogen-rich CSM only a few
months later \citep{2015ApJ...815..120M}.
 
In addition to the examples of interaction with dense CSM described above, indications of very short-lived interaction with 
CSM have been discovered through ``flash spectroscopy'' of very young SNe of various types
\citep[e.g.,][]{2014Natur.509..471G,2015ApJ...806..213S,2016ApJ...818....3K}.
These examples exist on a continuum of CSM densities with the strongly interacting events described above; they require much less CSM
and their observables at peak brightness generally align with those of ``normal'' events.

Here we present the results of our observational campaign to study
\thisSN, a remarkable and very well-monitored SN~Ibn.
It was discovered in NGC~2388 by the Lick Observatory Supernova Search (LOSS)
with the 0.76\,m Katzman Automatic Imaging Telescope \citep[KAIT;][]{2001ASPC..246..121F}
on 2015 Feb.\ 11 (all dates and times reported herein are UTC).
Note that, because the official International Astronomical Union name was not assigned until
November 2015 \citep{2015CBET.4164....1},
this event has also been discussed in the literature under the name PSN~J07285387+3349106.
\citet{2015ATel.7105....1O} classified it as a young SN~Ibn based upon
spectra obtained on Feb.\ 18, which showed a blue continuum and relatively narrow \ion{He}{1} emission
features \citep[see also][]{2015CBET.4164....1}. 
\citet{2015IBVS.6140....1T} present {\it BVRI} photometry of \thisSN\ starting Feb.\ 17,
showing that it has one of the fastest decline rates known 
(similar to those of SNe~2002bj, 2005ek, and 2010X) and is remarkably luminous (though
\thisSN\ is significantly obscured by an uncertain amount of dust in the host galaxy).
\citet{2015MNRAS.454.4293P} present additional photometry and low-resolution spectra of this event
and describe \thisSN\ within the context of SNe~Ibn
\citep[e.g.,][]{2007ApJ...657L.105F,2008MNRAS.389..113P,2016MNRAS.456..853P}.

In this paper we present photometric, spectroscopic, and spectropolarimetric
observations of \thisSN, including one epoch of relatively high-resolution Keck DEIMOS
spectroscopy ($R = \lambda/\delta \lambda \approx 5000$), enabling us to study the narrow-line features in detail.
We show that \thisSN\ is similar to several other SNe from the heterogeneous SN~Ibn class,
and that it shares many features with the rapid and luminous transients discovered in the Pan-STARRS1 (PS1) archives \citep{2014ApJ...794...23D},
those found in the SuperNova Legacy Survey (SNLS) archives \citep{2016ApJ...819...35A}, and
a few other rapidly fading SNe from the literature.  \thisSN\ offers valuable insights into the physics of the poorly observed class of rapidly fading SNe.

\section{Observations}
\label{sec:obs}

\subsection{Photometry}
\label{sec:phot}

 \begin{figure*}
\includegraphics[width=.48\textwidth]{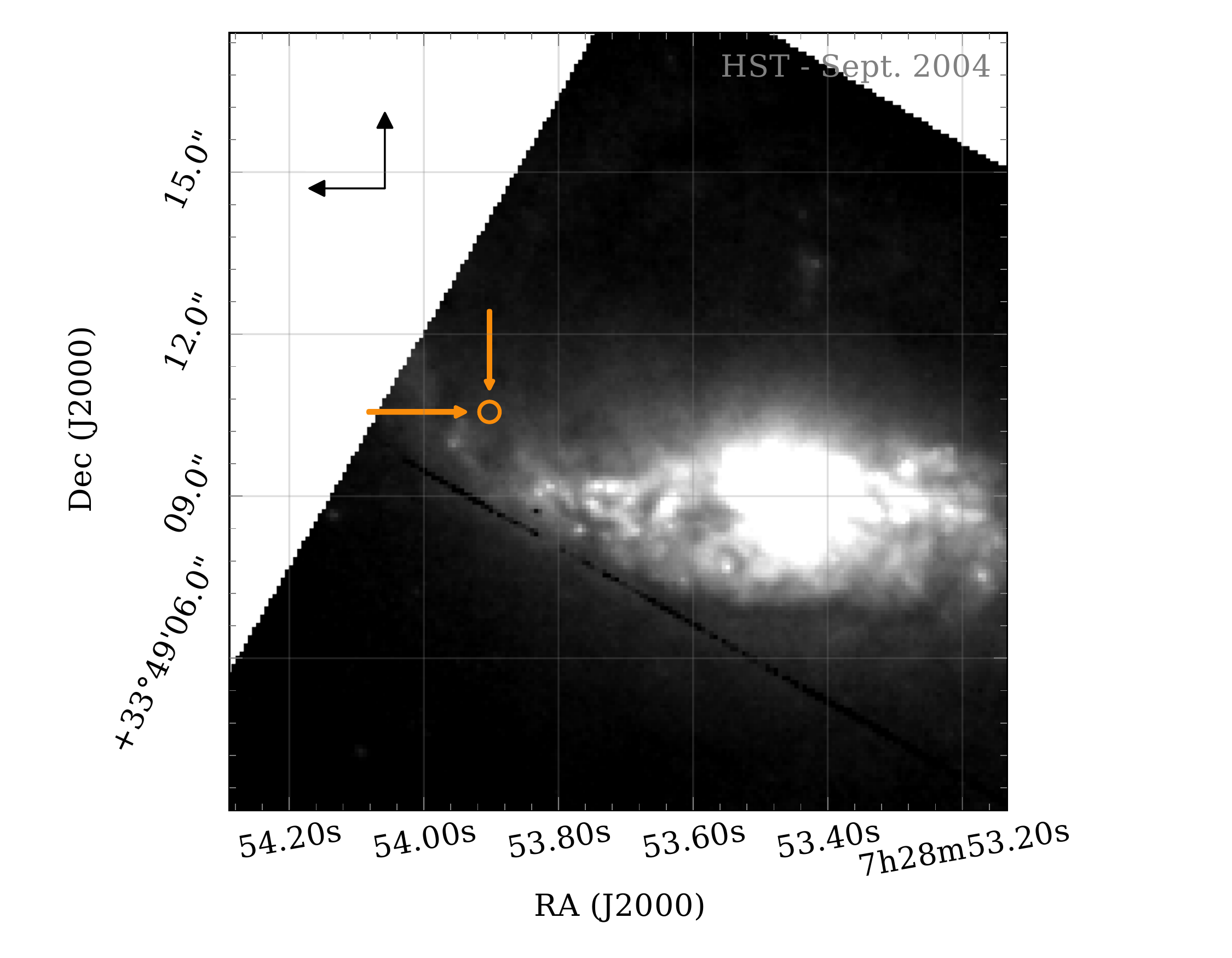}
\includegraphics[width=.48\textwidth]{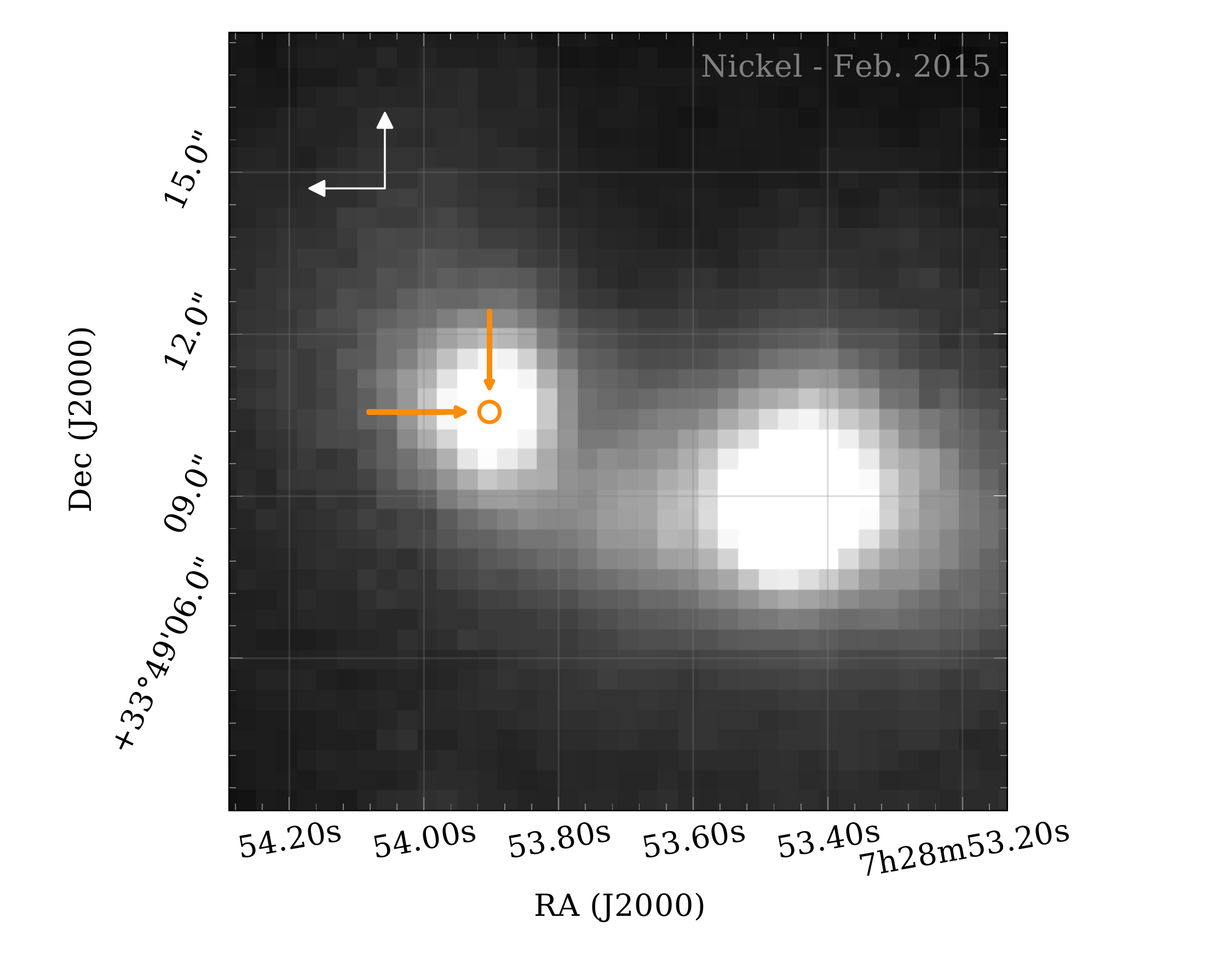} \\
\includegraphics[width=.48\textwidth]{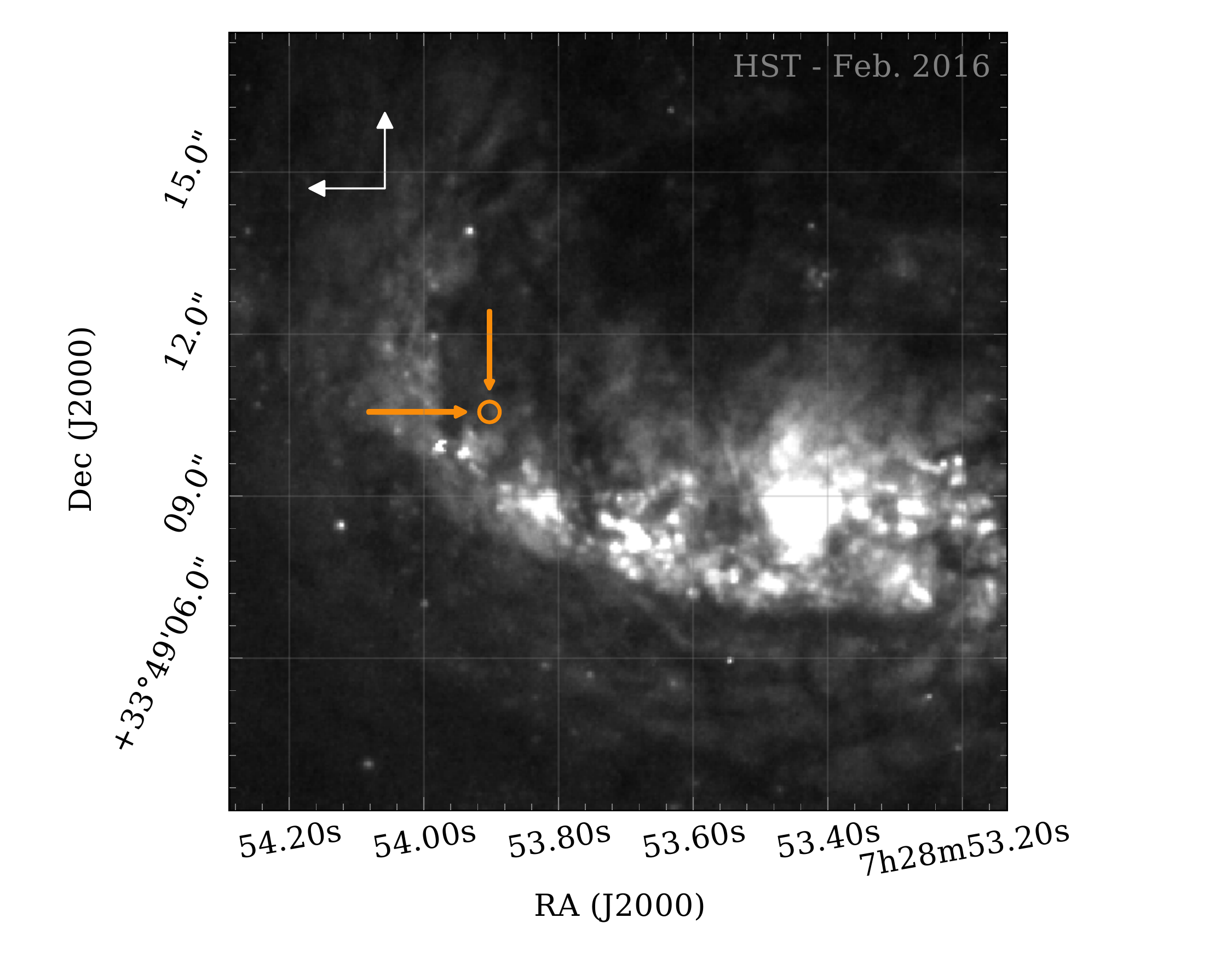}
\includegraphics[width=.48\textwidth]{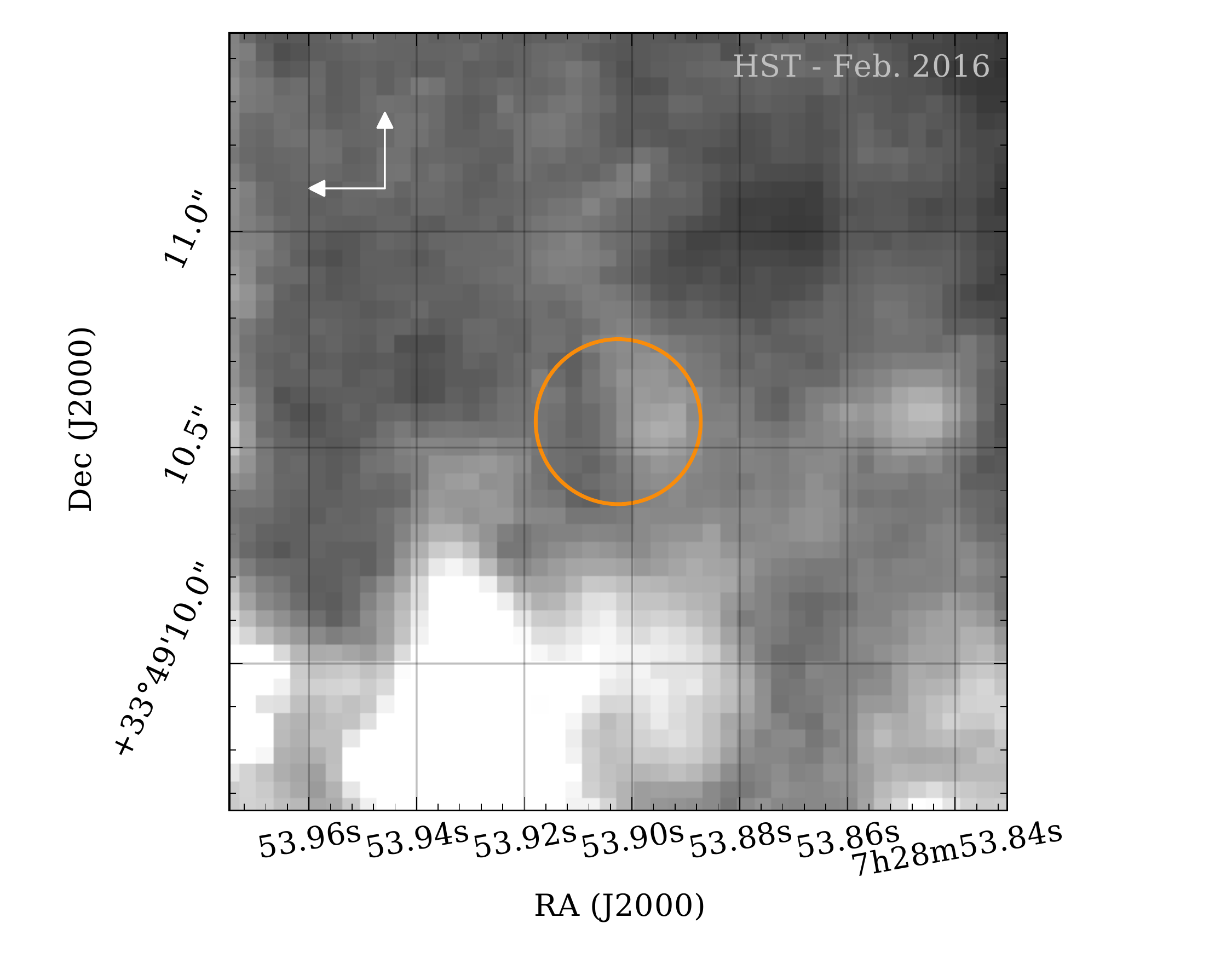}
 \caption{Top left: a pre-detection {\it HST} NICMOS image of NGC~2388 through the {\it F110W} filter.
                 Top right: a detection of \thisSN\ by the Lick Nickel 1\,m telescope through the $V$-band filter.
                 Bottom left: an {\it HST} WFC3 image of \thisSN's location through the {\it F555W} filter at $\sim1$\,yr post-explosion.
                 Bottom right: a zoom-in of \thisSN's location in the {\it HST} {\it F555W} filter.
                All images show the location of the SN with orange arrows, and the size of our 3$\sigma$ position error is marked with a circle. \label{fig:finder}}
 \end{figure*}

\thisSN\ was first detected by KAIT on Feb.\ 11.24 at $18.06 \pm 0.15$\,mag in an unfiltered
image.\footnote{The KAIT {\it clear} passband is similar to the $R$ band but broader. For more details and transformations to standard passbands, see \citet{1999AJ....118.2675R} and \citet{2003PASP..115..844L}.}
An unfiltered image taken the night before (Feb.\ 10.30) shows no source to a limit
of $\sim18.4$\,mag.
We began acquiring multiband photometry ({\it BVRI} and {\it clear}) starting
Feb.\ 14 with KAIT and the 1\,m Nickel telescope at Lick Observatory.
We used a set of 50 Nickel and KAIT images with strong detections of \thisSN\ and with
good astrometric solutions for the field to calculate an updated position:
$\alpha = 07^h28^m53.90^s$, $\delta = +33^{\degree}49'10.56''$ (J2000), offset from the centre
of the galaxy by $\sim6''$.  We believe this position to 
be accurate within $0.15''$ or better --- the positions we measured for the SN exhibit
a scatter of $0.09''$ in both right ascension and declination across 50 images.

\citet{2010ApJS..190..418G} describe our photometric observing program at Lick in detail,
along with our KAIT and Nickel image-reduction pipeline. Point-spread-function (PSF) photometry
was performed using DAOPHOT \citep{1987PASP...99..191S} from the IDL Astronomy
User's Library.\footnote{\url{http://idlastro.gsfc.nasa.gov/}}
Instrumental magnitudes were calibrated to several nearby
stars from the Sloan Digital Sky Survey, transformed into the
Landolt system using the empirical prescription
presented by Robert Lupton.\footnote{\url{http://www.sdss.org/dr7/algorithms/sdssUBVRITransform.html}}

We measure \thisSN's date of peak of brightness in each passband of our Nickel+KAIT photometry by fitting 
a low-order polynomial to the light curve within the first week (first 10 days for the $I$ band)
and use Monte Carlo Markov Chain (MCMC) methods to estimate our uncertainties.  In Modified Julian Day (MJD), we
find $t^{\rm max}_{V} = 57071.1 \pm 0.2$, $t^{\rm max}_{R} = 57071.6 \pm 0.5$, and 
$t^{\rm max}_{I} = 57072.4 \pm 0.2$ (our data do not constrain the $B$-band peak well).  
Throughout our analysis we present phases relative to the $V$-band peak.
NGC~2388 is at a redshift of $z_{\rm host} = 0.013790 \pm 0.000017$ \citep[NED;][]{1991rc3..book.....D},
which (assuming cosmological parameters H$_0 = 71$\,\kms\,Mpc$^{-1}$, $\Omega_m = 0.27$,
and $\Omega_{\Lambda} = 0.73$) translates into a luminosity distance of 58.9\,Mpc \citep{2006PASP..118.1711W}
and a distance modulus of $33.85$\,mag that we adopt for all absolute-magnitude corrections.

NGC~2388 and the SN site (pre-explosion) were imaged on 2004 Sep. 10 with 
the {\it Hubble Space Telescope (HST)}  and the Near-Infrared Camera and Multi-Object Spectrometer (NICMOS)
with the NIC2 aperture (scale $0{\farcs}076$ pixel$^{-1}$) in bands {\it F110W}, {\it F160W}, {\it F187N}, and {\it F190N}.
We determined the SN position in each band's mosaic data products based on the absolute SN position and the world coordinate system of the image ---
no stellar object was detected at this position in any of the mosaics. We quantified these nondetections in the {\it F110W} and {\it F160W} mosaics
using DAOPHOT by  assuming a PSF
constructed from the brightest isolated star in the mosaics and inserting an artificial star at the SN position.
The artificial star was measured with ALLSTAR within DAOPHOT using photometric calibrations established from the
online cookbook\footnote{\url{www.stsci.edu/hst/nicmos/performance/photometry/cookbook.html}; corrections to infinite aperture and zeropoints at zero magnitude were obtained from \url{www.stsci.edu/hst/nicmos/performance/photometry/postncs_keywords.html}.} and with parameters appropriate for NIC2,
and then reduced in luminosity until it was detected at a signal-to-noise ratio (S/N) of $\sim 3$.
The corresponding upper limits are $> 26.0$ and $> 25.2$\,mag in {\it F110W} and {\it F160W}, respectively.

Under our Cycle 23 Snapshot program with {\it HST}'s Wide Field Camera 3 (WFC3; GO-14149, PI Filippenko),
we obtained images of the SN location on 2016 Feb.\,14 (1\,yr post explosion)
through the {\it F555W} (710\,s) and {\it F814W} (780\,s) filters (scale $0{\farcs}04$ pixel$^{-1}$).
Figure~\ref{fig:finder} shows a Nickel detection of \thisSN\ alongside the pre-explosion NICMOS {\it F110W} image
and the 1\,yr WFC3 {\it F555W} image of the SN location.
We find that \thisSN\ exploded on the trailing edge of NGC~2388's spiral arm, in a region with several
probable dust lanes and a generally clumpy appearance.  One of those clumps falls within the
3$\sigma$ position error circle in both the {\it F555W} and {\it F814W} images and is not detected
in the pre-explosion NICMOS images.  However, this clump appears to be
extended and smoothly connected with other bright regions, not point-like, and we therefore
attribute it to the host galaxy and not to \thisSN.

\begin{figure}
\includegraphics[width=\columnwidth]{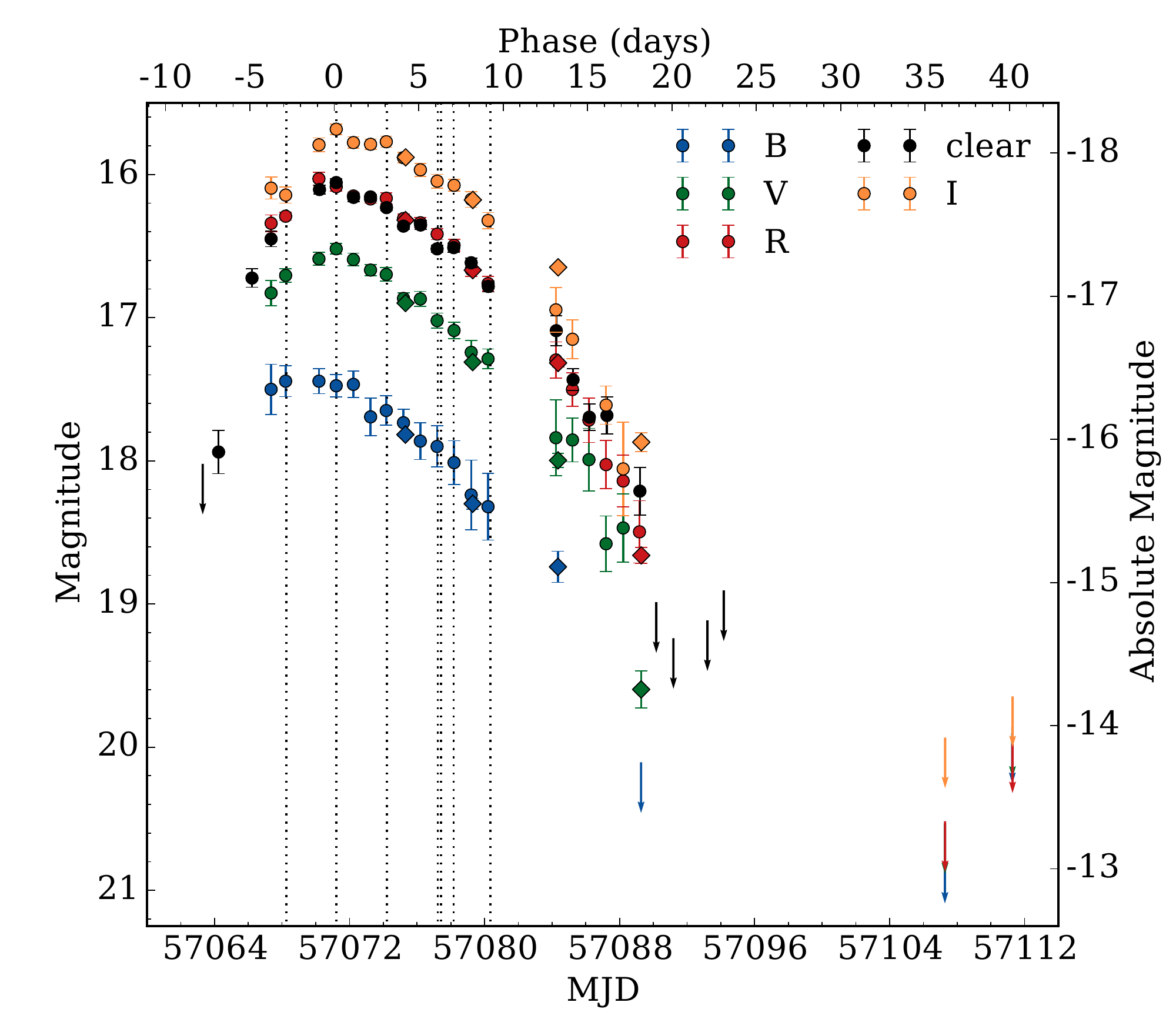}
 \caption{The {\it BVRI} and {\it clear} passband light curves of \thisSN\ 
    after correcting for MW dust absorption but not for the 
    host absorption.
    KAIT data are indicated by circles and Nickel data by diamonds.
    The dates of our spectral observations are marked by dotted vertical lines.
    The phase relative to the $V$-band peak
    is shown along the top while the absolute magnitude is shown on 
    the right.  Because these absolute magnitudes have not been corrected for
    significant but uncertain dust absorption arising within the host galaxy, they can be taken as 
    lower limits on the luminosity of \thisSN.
     \label{fig:phot}}
 \end{figure}

Table~\ref{tab:phot} presents our photometry of \thisSN\ before applying any dust reddening corrections,
and Figure~\ref{fig:phot} shows the light curves after correcting for Milky Way (MW) dust absorption.
The data are also available for download from
the Berkeley SuperNova DataBase \citep[SNDB;][]{2012MNRAS.425.1789S}.\footnote{\url{http://heracles.astro.berkeley.edu/sndb/}}
 
\begin{table*}
\caption{KAIT and Nickel Photometry of \thisSN}
\label{tab:phot}
\begin{minipage}{0.9\linewidth}
\centering
\begin{tabular}{ l l c c c c c c }
\hline
 UT Date &  MJD  &  {\it B (1$\sigma$)} & {\it V (1$\sigma$)} & {\it R (1$\sigma$)} & {\it I (1$\sigma$)} & {\it clear (1$\sigma$)} & Instrument \\
\hline
2015-02-10 & 57063.30 &  - &  - &  - &  - &  $>$18.4 &  KAIT \\ 
2015-02-11 & 57064.24 &  - &  - &  - &  - &  18.06 (0.15) &  KAIT \\ 
2015-02-13 & 57066.22 &  - &  - &  - &  - &  16.85 (0.06) &  KAIT \\ 
2015-02-14 & 57067.35 &  17.71 (0.17) &  16.98 (0.09) &  16.46 (0.06) &  16.18 (0.08) &  16.57 (0.05) &  KAIT \\ 
2015-02-15 & 57068.22 &  17.65 (0.11) &  16.86 (0.05) &  16.41 (0.03) &  16.23 (0.06) &  - &  KAIT \\ 
2015-02-17 & 57070.19 &  17.65 (0.09) &  16.74 (0.04) &  16.15 (0.05) &  15.88 (0.05) &  16.23 (0.03) &  KAIT \\ 
2015-02-18 & 57071.21 &  17.69 (0.08) &  16.67 (0.04) &  16.21 (0.03) &  15.77 (0.04) &  16.18 (0.03) &  KAIT \\ 
2015-02-19 & 57072.23 &  17.68 (0.09) &  16.75 (0.04) &  16.27 (0.03) &  15.87 (0.04) &  16.28 (0.03) &  KAIT \\ 
2015-02-20 & 57073.24 &  17.90 (0.13) &  16.82 (0.04) &  16.29 (0.03) &  15.88 (0.04) &  16.28 (0.02) &  KAIT \\ 
2015-02-21 & 57074.18 &  17.86 (0.10) &  16.85 (0.05) &  16.29 (0.04) &  15.86 (0.04) &  16.35 (0.02) &  KAIT \\ 
2015-02-22 & 57075.20 &  17.94 (0.09) &  17.02 (0.04) &  16.43 (0.03) &  15.97 (0.04) &  16.48 (0.02) &  KAIT \\ 
2015-02-22 & 57075.32 &  18.03 (0.02) &  17.05 (0.01) &  16.44 (0.01) &  15.97 (0.01) &  - &  Nickel \\ 
2015-02-23 & 57076.20 &  18.07 (0.13) &  17.02 (0.05) &  16.46 (0.04) &  16.06 (0.04) &  16.48 (0.03) &  KAIT \\ 
2015-02-24 & 57077.19 &  18.11 (0.14) &  17.18 (0.05) &  16.54 (0.04) &  16.13 (0.05) &  16.64 (0.02) &  KAIT \\ 
2015-02-25 & 57078.19 &  18.22 (0.15) &  17.24 (0.06) &  16.62 (0.04) &  16.16 (0.04) &  16.63 (0.03) &  KAIT \\ 
2015-02-26 & 57079.21 &  18.45 (0.24) &  17.40 (0.08) &  16.79 (0.05) &  16.26 (0.06) &  16.74 (0.03) &  KAIT \\ 
2015-02-26 & 57079.30 &  18.51 (0.04) &  17.47 (0.02) &  16.79 (0.01) &  16.27 (0.02) &  - &  Nickel \\ 
2015-02-27 & 57080.21 &  18.53 (0.23) &  17.44 (0.07) &  16.89 (0.05) &  16.41 (0.06) &  16.91 (0.04) &  KAIT \\ 
2015-03-03 & 57084.23 &  - &  17.99 (0.27) &  17.42 (0.13) &  17.03 (0.16) &  17.21 (0.10) &  KAIT \\ 
2015-03-03 & 57084.35 &  18.95 (0.11) &  18.15 (0.05) &  17.44 (0.03) &  16.74 (0.02) &  - &  Nickel \\ 
2015-03-04 & 57085.22 &  - &  18.01 (0.15) &  17.63 (0.12) &  17.24 (0.13) &  17.56 (0.08) &  KAIT \\ 
2015-03-05 & 57086.19 &  - &  18.15 (0.22) &  17.84 (0.16) &  - &  17.82 (0.09) &  KAIT \\ 
2015-03-06 & 57087.20 &  - &  18.73 (0.19) &  18.15 (0.17) &  17.70 (0.13) &  - &  KAIT \\ 
2015-03-06 & 57087.25 &  - &  - &  - &  - &  17.81 (0.13) &  KAIT \\ 
2015-03-07 & 57088.21 &  - &  18.62 (0.24) &  18.26 (0.18) &  18.14 (0.33) &  - &  KAIT \\ 
2015-03-08 & 57089.19 &  - &  - &  18.62 (0.22) &  - &  18.33 (0.17) &  KAIT \\ 
2015-03-08 & 57089.28 &  $>$20.5 &  19.75 (0.13) &  18.78 (0.06) &  17.96 (0.07) &  - &  Nickel \\ 
2015-03-09 & 57090.17 &  - &  - &  - &  - &  $>$19.3 &  KAIT \\ 
2015-03-10 & 57091.19 &  - &  - &  - &  - &  $>$19.6 &  KAIT \\ 
2015-03-12 & 57093.20 &  - &  - &  - &  - &  $>$19.5 &  KAIT \\ 
2015-03-13 & 57094.17 &  - &  - &  - &  - &  $>$19.3 &  KAIT \\ 
2015-03-26 & 57107.28 &  $>$21.1 &  $>$20.9 &  $>$20.9 &  $>$20.3 &  - &  Nickel \\ 
2015-03-30 & 57111.28 &  $>$20.3 &  $>$20.2 &  $>$20.3 &  $>$20.0 &  - &  Nickel \\ 

\hline
\end{tabular}
\end{minipage}
\end{table*}

\subsection{Spectroscopy}
\label{sec:spec_data}

We began a spectral monitoring campaign of \thisSN\ on 2015 Feb. 15, obtaining six spectra with the
Kast double spectrograph \citep{kast} on the Shane 3\,m telescope at Lick Observatory
and one spectrum with the DEIMOS spectrograph \citep{2003SPIE.4841.1657F} on the Keck-II 10\,m telescope.
Details of our spectral observations are listed in Table~\ref{tab:spec},
and all reductions and calibrations were performed with standard tools and methods, including IRAF routines and custom Python and IDL
codes\footnote{\url{https://github.com/ishivvers/TheKastShiv}} \citep[e.g.,][]{2000AJ....120.1487M,PER-GRA:2007,2012MNRAS.425.1789S}.

All spectra were taken at or near the parallactic angle \citep{1982PASP...94..715F},
and the amount of galaxy flux falling into the slit varied as a function of seeing and the 
slit orientation on the sky.
For the spectra which included a large amount of host-galaxy flux (Feb.\ 18, 21, and 24),
we extract a spectrum from a region of the galaxy that does not include any \thisSN\ flux,
perform a median-filter smoothing to obtain the galaxy continuum, and then subtract it from our spectra,
after determining a best-fit galaxy scaling coefficient by comparing synthetic photometry to the
multiband photometry observed the same night.  We do not attempt to remove the narrow (unresolved) galaxy emission features,
so our spectra show varying degrees of narrow-line contamination from the host (including H$\alpha$, H$\beta$, and [\ion{S}{2}]).

 \begin{figure*}
 \includegraphics[width=\textwidth]{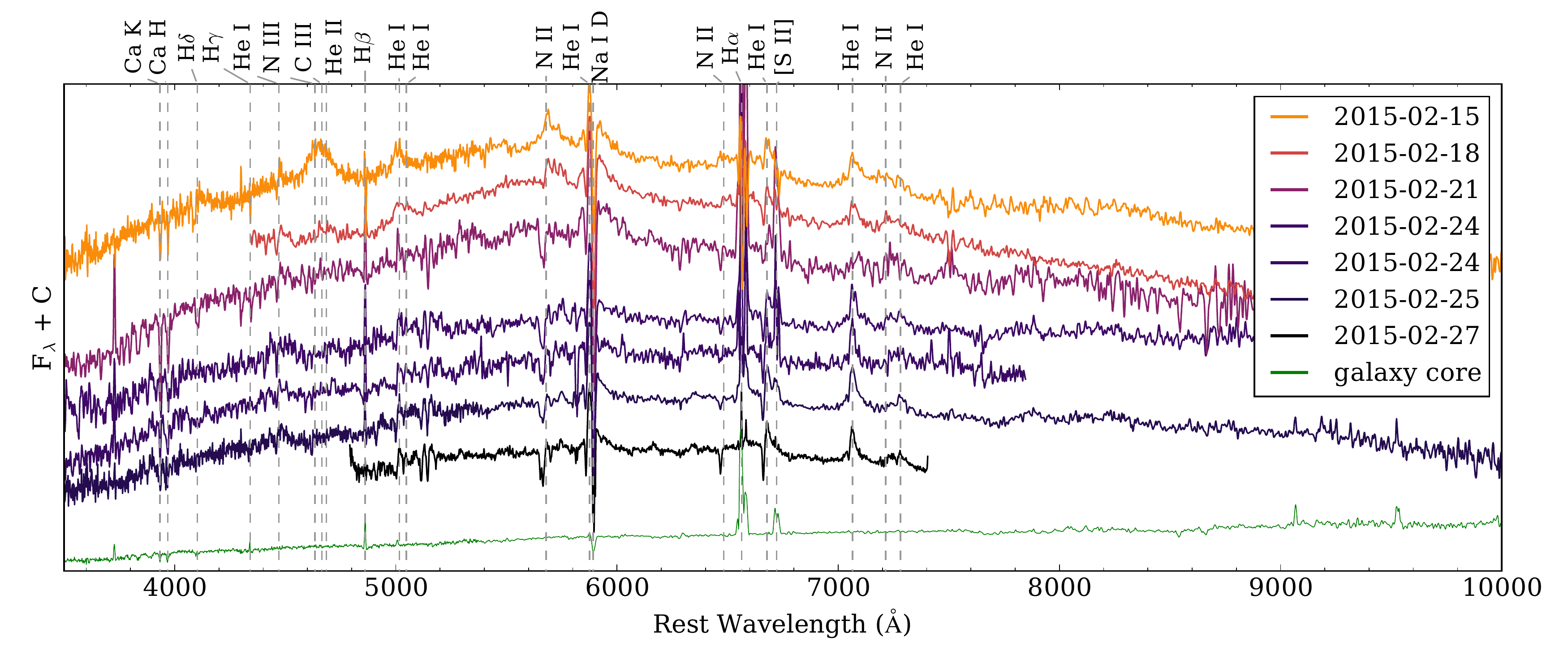}
 \caption{The spectral sequence of \thisSN.  
    This figure shows the spectral evolution after dereddening to
    correct for MW dust absorption but not for the host-galaxy absorption.
    On the bottom, we show an extraction of NGC~2388's nucleus, taken from a 
    Kast spectrum obtained on Feb.\ 15.
    Contamination from the host galaxy's strong emission lines is apparent in some spectra;
    we indicate H$\alpha$, H$\beta$, and [\ion{S}{2}]\,$\lambda\lambda6716$, 6731, as well as
    the host's \ion{Na}{1}\,D absorption.
     \label{fig:spec}}
 \end{figure*}

We renormalise each spectrum to match our template-subtracted $V$ or $R$-band photometry
(depending on the spectral wavelength coverage) after linearly interpolating the photometry to the time of
spectral observation.  Figure~\ref{fig:spec} shows our total-flux spectra of \thisSN.
All spectra are available to download from the Berkeley SNDB
and the Weizmann Interactive Supernova Data REPository
\citep[WiseREP;][]{2012PASP..124..668Y}.\footnote{\url{http://wiserep.weizmann.ac.il/}}

Using narrow H$\alpha$ host-galaxy emission in our DEIMOS spectrum, we measure a line-of-sight redshift
at the SN's location: $z_{\rm SN} = 0.013161 \pm 0.000005$, a difference of $\sim200$\,\kms\ from
the published redshift of the host galaxy --- consistent
with the SN's position on the approaching arm of this spiral galaxy.

\subsection{Spectropolarimetry}
\label{sec:specpol}

Three epochs of spectropolarimetry were obtained of \thisSN\ near peak brightness utilising
the dual-beam polarimetry mode of the Lick 3\,m Kast spectrograph. The orientation of the slit on the sky was always set to a position angle of 
$180^{\circ}$ (i.e., aligned north-south), and exposures of 900\,s were obtained at each of four waveplate positions 
($0\fdg0$, $45\fdg0$, $22\fdg5$, and $67\fdg5$). On each night, several waveplate sequences were performed and coadded.
Flatfield and arc-lamp spectra were obtained immediately after each sequence without moving the telescope. 

For polarimetric calibrations, the low-polarization standard stars BD+32$^{\circ}$3739 and BD+05$^{\circ}$2618 were observed 
to verify the low instrumental polarization of the Kast spectrograph. We constrained the average fractional Stokes $Q$ and $U$ 
values to $<0.1$\%. By observing the above unpolarized standard stars through a 100\% polarizing filter we determined that 
the polarimetric response is so close to 100\% that no correction was necessary.  Finally, we obtained the instrumental polarization 
position-angle curve and used it to correct the data.
We observed the high-polarization stars HD\,19820, BD\,+59$^{\circ}$389, 
and V1Cyg12 to obtain the zeropoint of the polarization position angle on the sky ($\theta$) and to determine the accuracy of 
polarimetric measurements, which were generally consistent with previously published values within $\Delta P<0.05$\% and 
$\Delta \theta<1^{\circ}$. All spectropolarimetric reductions and calculations were performed using the methods described
by \citet[and references therein]{2015MNRAS.453.4467M}, and we define the polarimetric parameters in the same manner as those authors
(Stokes parameters $q$ and $u$, debiased polarization $P$, and sky position angle $\theta$). 

To probe the Galactic component of interstellar polarization (ISP) along the line of sight toward \thisSN, we used the 
method of spectroscopic parallax to select three stars within $1^\circ$ of separation from the source and measured
their integrated $V$-band polarization ($P$) and position angle ($\theta$). For
HD\,58221 we obtained $P = 0.28\%\pm$0.01\%, $\theta = 13.0^{\circ}\pm0.6^{\circ}$; for
HD\,58726, $P = 0.50\%\pm$0.01\%, $\theta = 24.8^{\circ}\pm0.3^{\circ}$;
and for HD\,59291, $P = 0.24\%\pm$0.01\%, $\theta = 16.3^{\circ}\pm0.7^{\circ}$ (uncertainties are statistical).
We interpret the significantly higher value of $P$ for  HD\,58726 (A0~V spectral type) 
as being caused by some intrinsic polarization for that star, so we used the average values of the other two stars (0.25\%, 14.3$^{\circ}$) to 
calculate the associated Serkowski-Whittet form \citep{1975ApJ...196..261S,1992ApJ...386..562W}, assuming a total-to-selective extinction ratio of 
$R_V=3.1$ and that the polarization peaks at 5500\,{\AA}, values appropriate for the MW. The resulting curve was subtracted 
from the \thisSN\ data to remove the small but nonnegligible Galactic component of ISP. 
The above observations (along with calibration observations of the low-polarization standard star HD\,14069 and the high-polarization 
standards HD\,245310 and HD\,25443) were taken with the Kast spectrograph on 2015 Sep.\ 21.

\begin{table*}
\caption{Journal of Spectroscopic Observations}
\label{tab:spec}
\begin{minipage}{0.9\linewidth}
\centering 
\begin{tabular}{ l | c c c c c c c }
\hline
UT Date & Type\footnote{F: total flux; S: spectropolarimetry.} & Tel./Instr. & Wavelength &  Resolution &
  Exp.  &  Observer\footnote{\label{obsnote}Observers and data reducers are indicated with their initials. IS: Isaac Shivvers; WZ: WeiKang Zheng; JM: Jon Mauerhan; MG: Melissa Graham; PK: Patrick Kelly; IK: Io Kleiser; JS: Jeffrey Silveman.} & 
Reducer\footref{obsnote} \\
   &   &   &  (\AA) & (\AA)  &  (s)  &   &    \\
\hline
2015-02-15.256 & F & Shane/Kast  & 3440--10,870 & 2   & 1200 & IS & IS \\
2015-02-18.211 & F & Shane/Kast  & 4400--9880   & 2   & 2400 & WZ & IS \\
2015-02-21.207 & F & Shane/Kast  & 3500--10,500 & 2   & 1200 & WZ & IS \\
2015-02-21.313 & S & Shane/Kast  & 4500--10,000 & 16 & $4 \times 4 \times 900$\footnote{All spectropolarimetry was obtained by rotating multiple times through four waveplate positions; see \S\ref{sec:specpol}.} & WZ & JM \\
2015-02-24.224 & F & Shane/Kast  & 3500--10,500 & 2   & 1200 & WZ & IS \\
2015-02-24.322 & S & Shane/Kast  & 4500--10,000 & 16 & $3 \times 4 \times 900^c$ & WZ & JM \\
2015-02-24.414 & F & Shane/Kast  & 3500--7875   & 1   & 1800 & WZ & IS \\
2015-02-25.171 & F & Shane/Kast  & 3500--10,500 & 2   & 1200 & MG & MG \\
2015-02-25.301 & S & Shane/Kast  & 4500--10,000 & 16 & $5 \times 4 \times 900^c$ & MG & JM \\
2015-02-27.34  & F & Keck/DEIMOS & 4850--7510   & 0.3 & 1200 & PK & IS \\
2010-03-022.17\footnote{A previously unpublished observation of SN~2010al; see Figure~\ref{fig:spec_comparisons}.} & F & Shane/Kast & 3450--10,790 & 3 & 2100 & IK & JS \\
\hline
\end{tabular}
\end{minipage}
\end{table*}

\section{Analysis}
\label{sec:analysis}

\subsection{Dust Corrections}
\label{sec:dust}

All observations of \thisSN\ are heavily affected by an obscuring screen of dust
present in the host galaxy, NGC~2388.  Because the treatment of
dust corrections affects much of the following discussion, we start
by describing our efforts to understand, characterise, and correct for the effects
of the host-galaxy dust. First, however, we correct for MW dust along the line of sight using the dust maps of 
\citet[$E(B-V) = 0.0498$\,mag;][]{2011ApJ...737..103S} and assuming $R_V=3.1$ \citep{1989ApJ...345..245C}.

\citet{2015ATel.7105....1O} note that a significant amount of host-galaxy reddening
is apparent in their classification spectrum, with an estimated $E(B-V) = 1.0$\,mag
based upon the \ion{Na}{1}\,D feature.
Our higher-resolution DEIMOS spectrum from Feb.\ 27 reveals complex 
\ion{Na}{1}\,D absorption near the host redshift (we examine the structure of this feature in \S\ref{sec:csm}).
We measure the equivalent width (EW) of the entire absorption complex to be
$9.1 \pm 0.15$\,\AA\ --- well outside the empirical relations of \citet{2012MNRAS.426.1465P} and similar
previous efforts, which show that the \ion{Na}{1}\,D lines saturate and lose their predictive power above a total EW
of $\sim2$\,\AA\ and $E(B-V) \approx 1.0$\,mag.  This indicates that the reddening toward \thisSN\ 
likely exhibits $E(B-V) > 1.0$\,mag, if the dust in NGC~2388 is similar to that in our MW.  

\citet{2015MNRAS.454.4293P} measure an \ion{Na}{1}\,D EW of $6.5 \pm 0.5$\,\AA\ from their spectra; we tentatively suggest
that the discrepancy between their value and ours is caused by different choices made when defining the local continuum.
Figure~\ref{fig:HeI} shows that the \ion{Na}{1}\,D components fall on the red wing of the \ion{He}{1} emission line, even at early
times.  However, if one instead defines the local continuum based upon the overall flux-density level (a reasonable choice if the emission-line wings
were not detected), one obtains \ion{Na}{1}\,D EW values similar to those of \citet{2015MNRAS.454.4293P}.

We next examined the 5780.5\,\AA\ diffuse interstellar band (DIB). This feature is one of the strongest
DIBs; it has long been known to correlate with extinction in the MW when the \ion{Na}{1}\,D 
feature is saturated
\citep[e.g.,][]{1995ARA&A..33...19H,2011ApJ...727...33F,2015MNRAS.452.3629L,2015MNRAS.447..545B}, and \citet{2013ApJ...779...38P}
show a clear correlation between this feature and the reddening toward 
SNe~Ia produced by host-galaxy dust. We do not detect this feature, and our spectra have insufficient
S/N to place strong constraints on the dust using our nondetection.
We also examined modeled estimates for the amount of dust obscuring the bulk of stars in the host
galaxy NGC~2388 \citep{2015A&A...577A..78P}, and we determined the same value at the location of the SN
(within $\sim1''$ on the sky) by measuring the Balmer ratio implied by the H$\alpha$
and H$\beta$ galaxy lines present in our DEIMOS spectrum \citep[e.g.,][]{1971MNRAS.153..471B}.
We estimate $E(B-V) \approx 0.56$ and $1.1$\,mag, respectively, toward the average star in NGC~2388's core and 
the average star near the explosion site of \thisSN.
Although this does not provide a measure of the line-of-sight reddening toward \thisSN\ specifically,
it does provide valuable context about the dust content of NGC~2388.

Additional information is available to us through the spectropolarimetry: as we describe in \S\ref{sec:specpolAnalysis}, 
the peak of the polarization spectrum ($\lambda_{\rm max}$) of NGC~2388's ISM is blueward of 4600\,\AA.  The shapes of the polarization law and the 
dust-extinction law are related within the MW \citep[e.g.,][]{1975ApJ...196..261S,1988ApJ...327..911C}.
Using $R_V = (-0.29 \pm 0.74) + (6.67 \pm 1.17)\,\lambda_{\rm max}$ from \citet[with $\lambda_{\rm max}$ in $\rm{\mu}$m]{1988ApJ...327..911C},
our measurement of $\lambda_{\rm max} < 0.46\,\rm{\mu}$m implies $R_V < 2.8 \pm 0.9$.
However, note that \citet{2015A&A...577A..53P} show the polarization properties of the host-galaxy dust
obscuring several well-studied and reddened SNe~Ia to be remarkably different from those of the dust in our MW.

 \begin{figure}
 \includegraphics[width=\columnwidth]{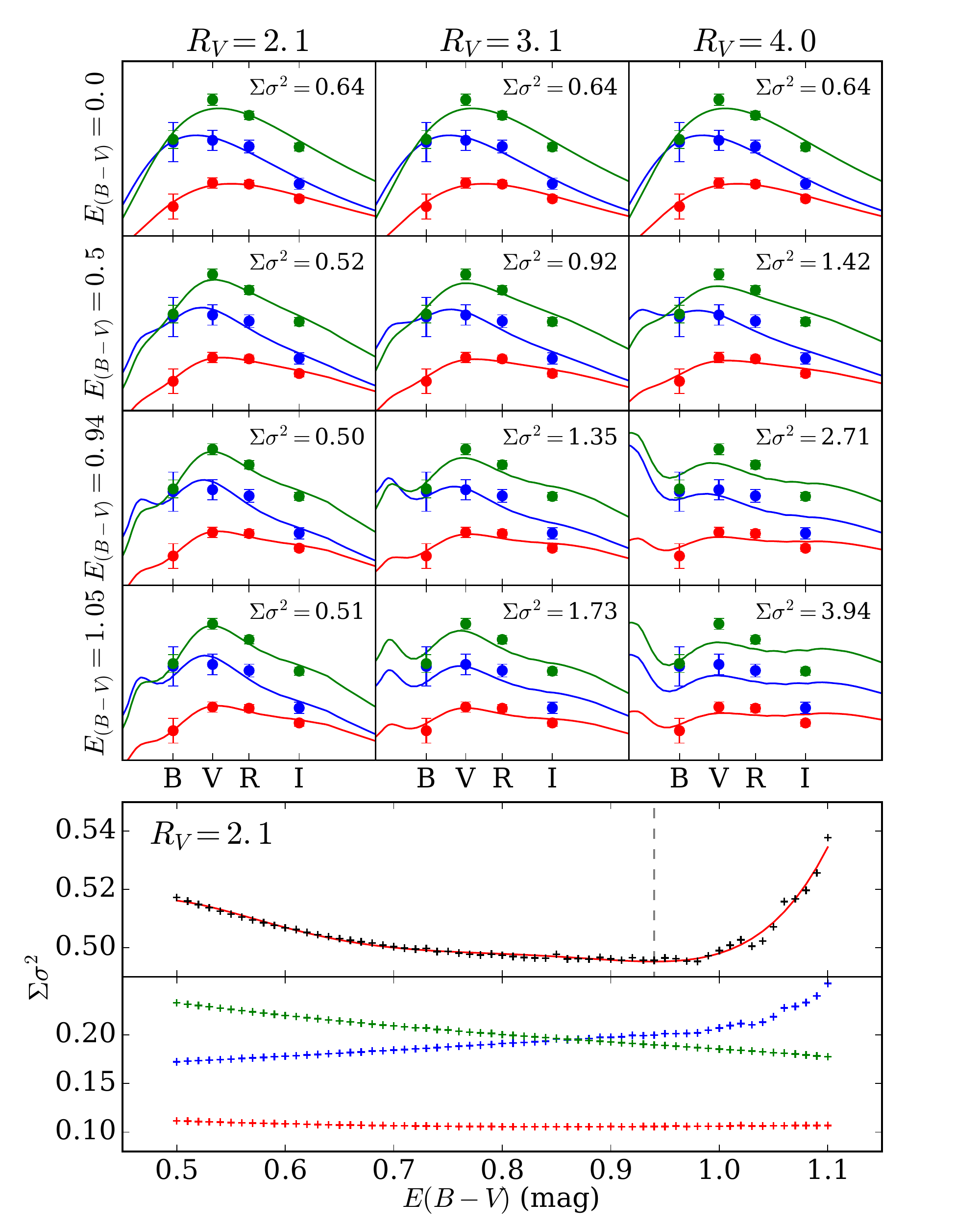}
 \caption{In the top panel, we fit reddened blackbodies to our photometry of \thisSN\ at three epochs 
    (blue, Feb.\ 14.35; green, Feb.\ 18.21; red, Feb.\ 27.21) after correcting for MW absorption
    and assuming various values for $R_V$ and $E(B-V)$ within NGC~2388.
    For many data points, the measurement errors are smaller than the plotted symbol.
    The weighted sum of squared differences ($\Sigma\sigma^2$) between observed photometric points and synthetic photometry of a blackbody
    is shown in the upper right of each plot; lower values indicate a better agreement across all epochs.
    In the bottom panel, we show $\Sigma\sigma^2$ as a function of $E(B-V)$ for $R_V = 2.1$ in black as well as $\Sigma\sigma^2$ broken down
    by individual epochs, coloured as above.
    $R_V = 2.1$ and $E(B-V) = 0.94^{+0.1}_{-0.4}$\,mag (marked with a dashed line in the bottom panel) are the preferred values.  
    In many panels a blue feature is introduced into the 
    reddened blackbody spectrum by the wavelength dependency of the MW's reddening law, but our spectral data provide no clear evidence for or against
    such a feature in the spectra of \thisSN.
\label{fig:bbody_phot}}
\end{figure}

Finally, we model \thisSN\ as a blackbody emitter obscured behind a simple screen of dust in the host galaxy and fit for the parameters of that dust
via comparisons with our multiband photometry at three epochs (pre-maximum, maximum, and post-maximum brightness).
Our spectra show that the emission from \thisSN\ is roughly that of a blackbody obscured by dust, at least at optical wavelengths
(unlike most SNe, which display spectra consisting of prominent and overlapping features with sometimes strong line blanketing),
and in \S\ref{sec:temp} we discuss why one would expect roughly blackbody emission from \thisSN.
We use PySynphot\footnote{\url{http://ssb.stsci.edu/pysynphot/docs/}} and
assume the dust is described by the empirical extinction law of \citet{1989ApJ...345..245C} --- by varying the parameters
$R_V$ and $E(B-V)$ we are able to explore
a wide variety of dust populations and search for the parameters that produce the best blackbody fits to our photometric data.

\begin{figure*}
 \includegraphics[width=\textwidth]{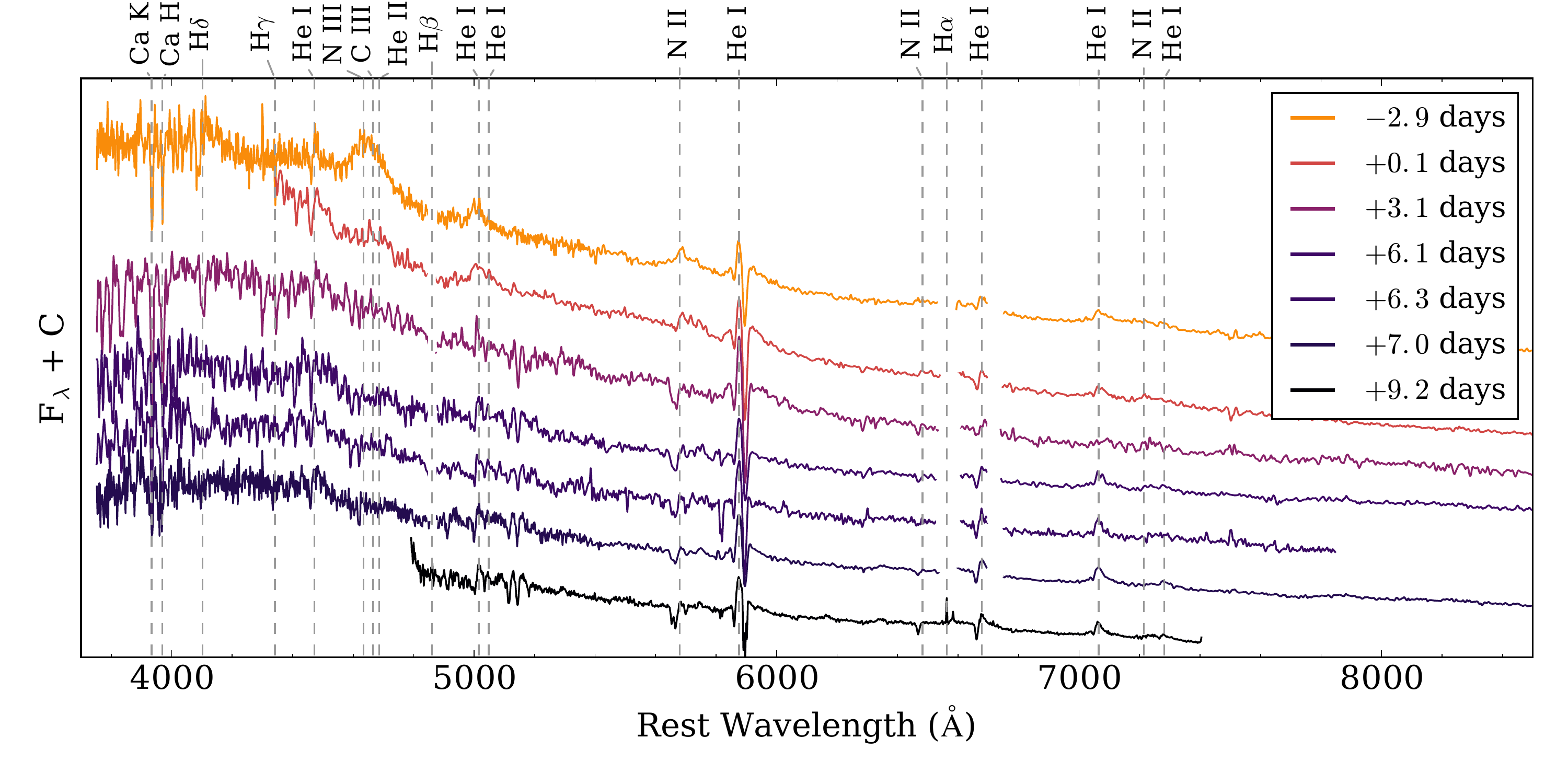}
 \caption{The spectral sequence of \thisSN\ after dereddening to 
    correct for MW dust absorption and the (uncertain) host absorption.  Phases are
    listed relative to the $V$-band peak brightness.
    We have masked regions of our low-resolution spectra where the host's narrow
    emission lines dominate, but we do not mask the \ion{Na}{1}\,D absorption features from the MW
    and the host.
     \label{fig:spec_correct}}
 \end{figure*}

\begin{figure*}
 \includegraphics[width=0.45\textwidth]{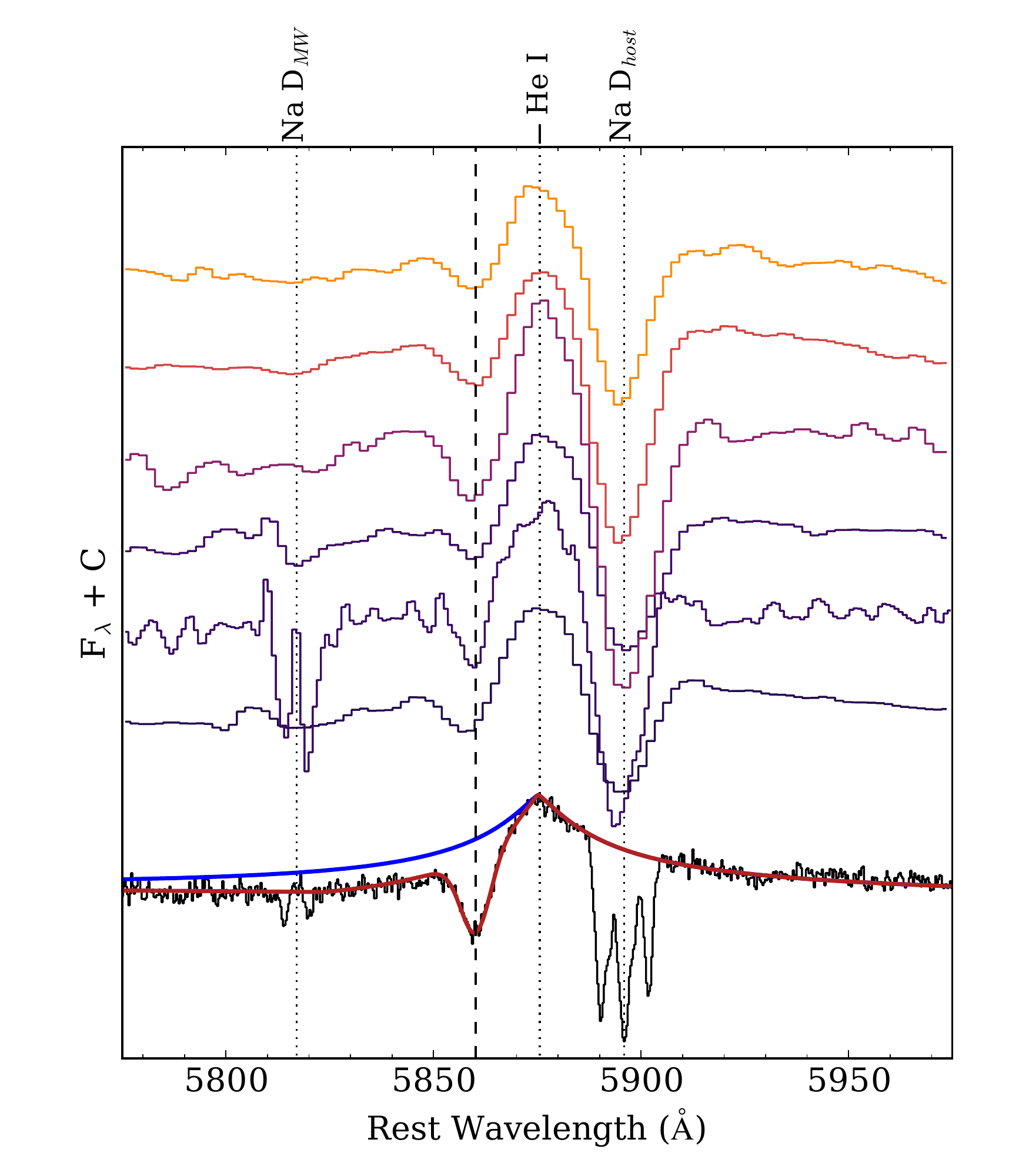}
 \includegraphics[width=0.45\textwidth]{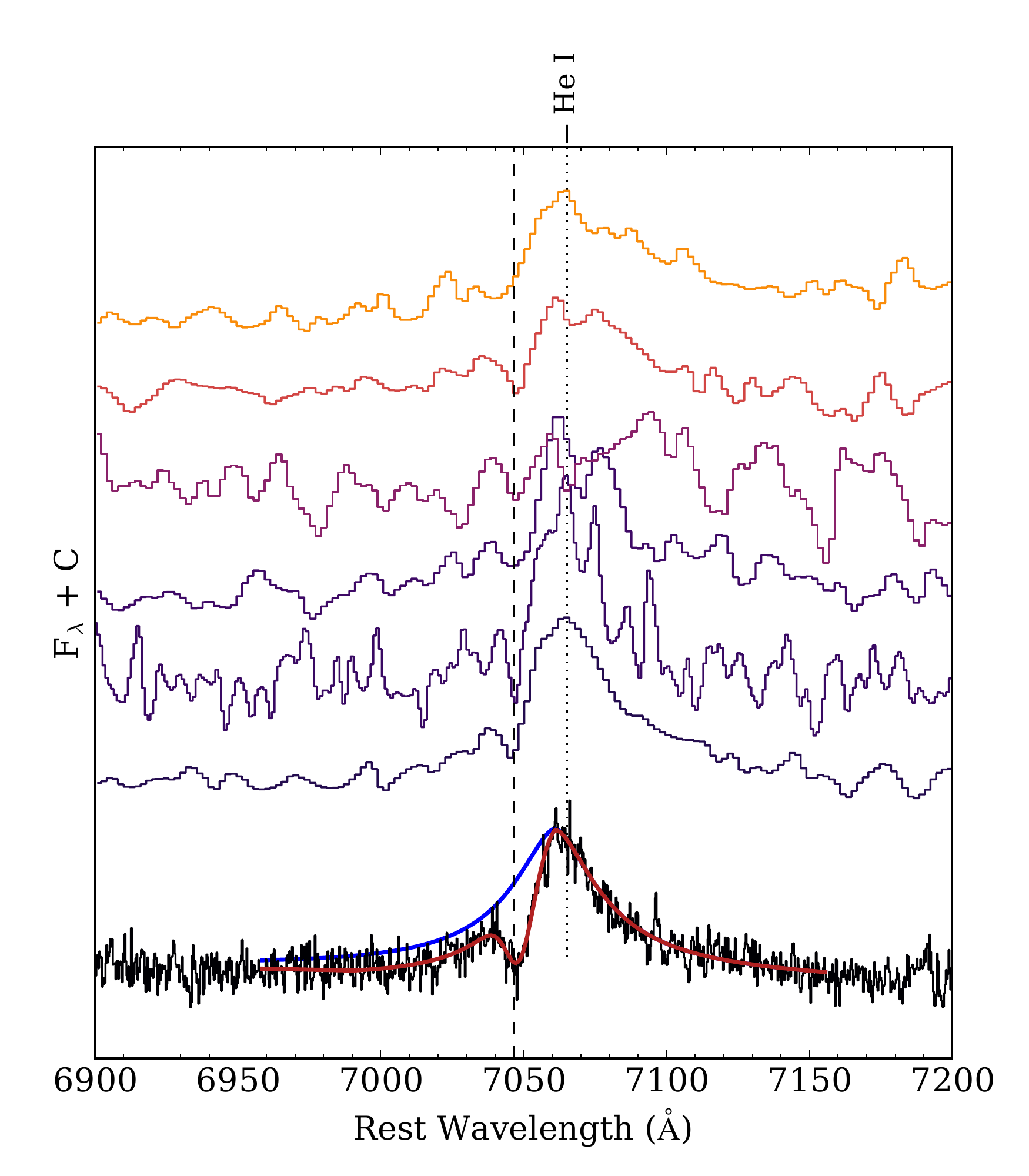}
 \caption{ The evolution of the \ion{He}{1}\,$\lambda$5876 (left) and $\lambda$7065 (right) lines. The spectra increase in time downward on the plot
    as in Figure~\ref{fig:spec}. The rest wavelength of each line is shown with a dotted line and the absorption component is marked
    at a blueshift of $v = 789$\,\kms\ (the mean blueshift of \ion{He}{1} lines measured from our DEIMOS spectrum).
    In the left panel we also mark the host and MW \ion{Na}{1}\,D absorption features
    (see \S\ref{sec:csm} for a discussion of the host galaxy's remarkable \ion{Na}{1}\,D absorption complex).
 \label{fig:HeI} }
 \end{figure*}

Figure~\ref{fig:bbody_phot} shows a comparison between synthetic photometry from PySynphot and the photometry
assuming four different values for $E(B-V)$ ranging from 0.0\,mag to 1.0\,mag and three different reddening laws parameterised
by $R_V$ from 2.1 to 4.0 (note that PySynphot only includes a few built-in options for $R_V$; we explore the most relevant of them here).
For every combination of $R_V$ and $E(B-V)$, we perform a weighted least-squares fit to determine the best-fit temperature and luminosity of the
source blackbody and we list the weighted sum of squared differences ($\Sigma\sigma^2$) for each.
Figure~\ref{fig:bbody_phot} shows that a nonzero amount of host-galaxy reddening correction makes the fits worse for $R_V = 3.1$ and 4.0 ---
in those cases the photometric data prefer a blackbody with no dust.
However, the spectra clearly show a nonthermal rolloff at blue wavelengths (see Figure~\ref{fig:spec}), and so
there must be a significant amount of dust absorption within NGC~2388.
If we instead adopt $R_V = 2.1$, a grid search over $E(B-V)$ values indicates a best fit at $E(B-V) \approx 0.94$\,mag.
As Figure~\ref{fig:spec_correct} shows, dereddening the spectra for this reddening law appears to correct the blue rolloff.
Figure~\ref{fig:bbody_phot} shows that $\Sigma\sigma^2$ increases above $E(B-V) = 1.0$ and below $E(B-V) = 0.6$\,mag, but
between those values $\Sigma\sigma^2$ is only weakly dependent upon $E(B-V)$.
We therefore use Figure~\ref{fig:bbody_phot} to estimate our error bars: $E(B-V) = 0.94^{+0.1}_{-0.4}$\,mag.
A similar approach was taken by \citet{2015MNRAS.454.4293P}, who compared the colour curves of \thisSN\ 
to the intrinsic colour curves of other SNe~Ibn to obtain $E(B-V)_{\rm tot} = 0.81 \pm 0.21$\,mag (assuming
$R_V = 3.1$).  However, as they note, the SN~Ibn subclass is remarkably heterogeneous, and it is not clear
whether the physics governing those colour curves is the same for all members.

Given the above discussion, we adopt $R_V = 2.1$ and $E(B-V) = 0.94^{+0.1}_{-0.4}$\,mag.
The large EWs measured from the \ion{Na}{1}\,D and the odd absorption-feature complex indicate that a simple analysis
of those features cannot be trusted. Studies of the integrated galaxy flux and the Balmer
decrement measured from our spectrum show that the bulk of NGC~2388's stellar mass is strongly obscured,
and our spectropolarimetric data indicate that \thisSN\ itself exploded behind a significant dust screen.
The low $\lambda_{\rm max}$ value in the polarization spectrum and our blackbody
fits to the light curves together prefer a relatively wavelength-dependent dust extinction law for NGC~2388 (i.e.,\ $R_V < 3.1$) and 
$E(B-V) = 0.94^{+0.1}_{-0.4}$\,mag --- we find this line of reasoning most convincing.
The above result is consistent with the $E(B-V) = 0.99 \pm 0.48$\,mag adopted by \citet{2015MNRAS.454.4293P}, though they assume $R_V = 3.1$.
Our determination of the dust properties toward \thisSN\ remains uncertain, so we consider a range of plausible reddening corrections
throughout the rest of this paper.  We find that the extinction toward \S\ref{sec:csm} does not appear to change over the course of our observations,
and we discuss and reject the possibility that a significant fraction of the extinction toward \thisSN\ 
arises from the SN's CSM rather than the host galaxy's ISM.

\subsection{Spectra}
\label{sec:spectra_analysis}

The optical spectra of \thisSN\ show a strong continuum overlain with narrow and intermediate-width emission features.
Though our data cover only 12 days of \thisSN's evolution, they range from a pre-maximum spectrum to
one taken after the SN had faded from peak by $\sim1$\,mag (see Figure~\ref{fig:phot}).  
No broad ejecta features became apparent in our spectra throughout that timespan
--- markedly different behaviour from that of the well-observed, rapidly fading 
SNe 2002bj, 2005ek, and 2010X \citep{2010Sci...327...58P,2013ApJ...774...58D,2010ApJ...723L..98K}
and similar to that of the interacting SN~2010jl \citep[e.g.,][]{2014ApJ...797..118F}.
Though broad features do not emerge, the continuum temperature shows significant evolution, the
narrow emission and absorption features evolve, and the relative ionisation states of the detected elements change.
While there are no strong H lines, 
multiple ionisation levels of helium, nitrogen, iron, and perhaps oxygen and carbon are detected, some in both emission and absorption and some in absorption only.
\citet{2015MNRAS.454.4293P} also present a series of spectral observations of \thisSN.  Their data are in good
agreement with ours and extend to later epochs, while our (higher-resolution) spectra illustrate new details not apparent previously.

\subsubsection{Evolving Line Profiles}
\label{sec:lineprofiles}

The most obvious features in our spectra are emission lines of \ion{He}{1}.  These lines 
are centred at the rest wavelength, exhibit a relatively broad base
with a full width near zero intensity (FWZI) of $\sim9000$--10,000\,\kms, and are overlain with P-Cygni 
absorption features blueshifted by $\sim745$\,\kms~and with widths of 400--500\,\kms.
The wavelengths and widths of these features do not significantly change during the $\sim12$ days covered by our spectra.
We find that the \ion{He}{1} line profiles can be parameterised by a ``modified Lorentzian'' emission component (where the exponent is allowed to vary)
with a suppressed blue wing and an overlain Gaussian absorption component, indicative of recombination lines broadened
via electron scattering within the CSM.  \thisSN's line profiles are very similar to those observed in many SNe~IIn but with a faster inferred CSM velocity
\citep[see, e.g., the spectra of SN~1998S;][]{2000ApJ...536..239L,2001A&A...367..506A,2001MNRAS.325..907F,2001MNRAS.326.1448C,2015ApJ...806..213S}. 
Figure~\ref{fig:HeI} shows our best fit to the \ion{He}{1}\,$\lambda$5876 line using this parameterisation.
We also show the unsuppressed Lorentzian feature to emphasise the asymmetry, which arises from the effects of Compton
scattering within a wind-like CSM \citep[e.g.,][]{1972ApJ...178..175A,1991A&A...247..455H,2001MNRAS.326.1448C}.
The magnitude of the effect scales with the expansion velocity, explaining the strongly asymmetric emission lines of \thisSN\ compared to 
those observed in most SNe~IIn.

\begin{figure}
 \includegraphics[width=\columnwidth]{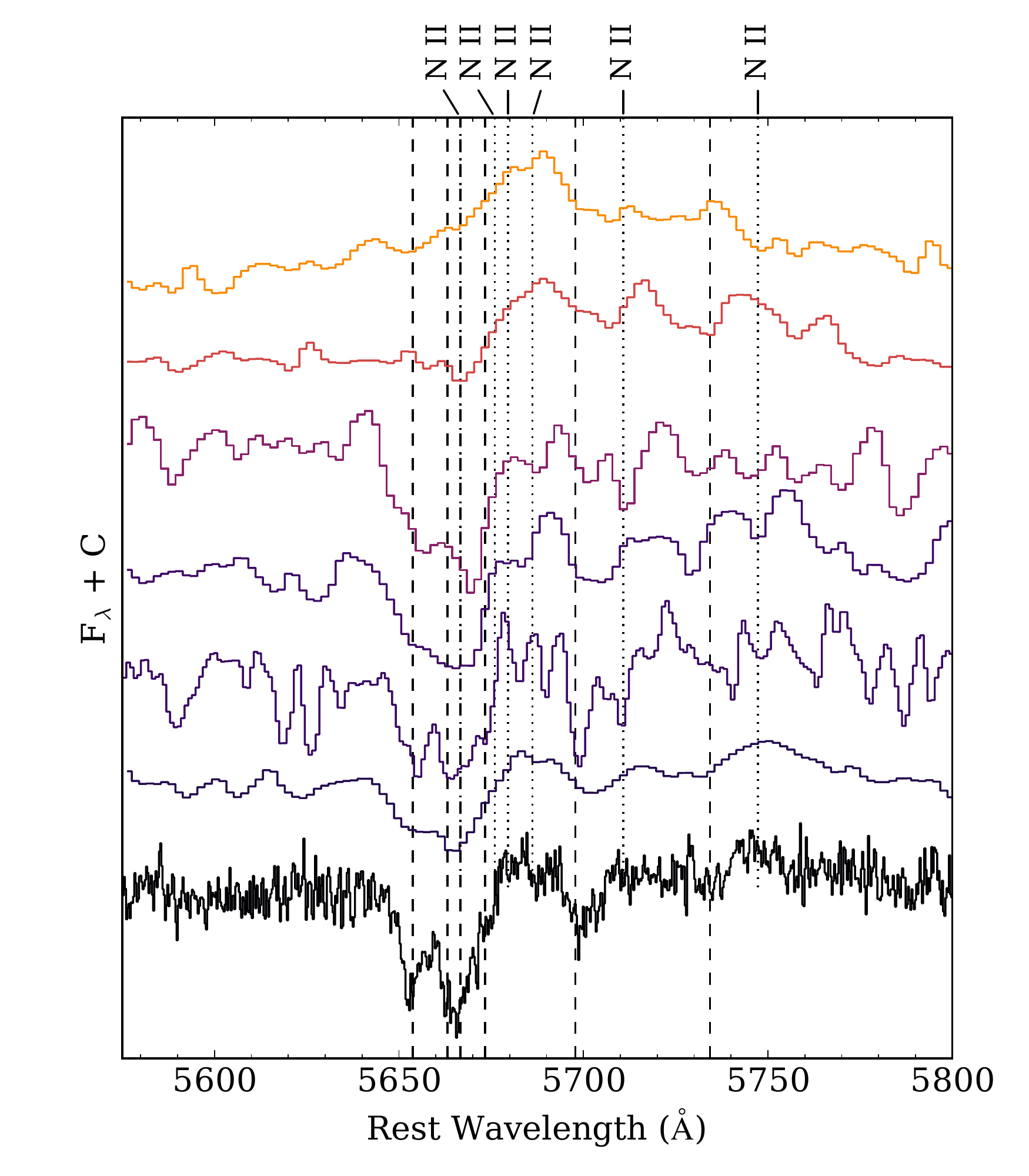}
 \caption{The evolution of the \ion{N}{2} lines.  The spectra increase in time downward
    as in Figure~\ref{fig:spec}.  Identified \ion{N}{2} lines are marked at rest using dotted lines and with
    a blueshift of $v = 680$\,\kms\ using dashed lines (the mean blueshift of \ion{N}{2} lines measured from our DEIMOS spectrum). 
   The emission features marginally detected at $v = 0$\,\kms~in our earliest spectra evolve into blueshifted
   absorption lines post-maximum brightness.
 \label{fig:NaII} }
 \end{figure}

After \ion{He}{1} (and host-galaxy H$\alpha$ and [\ion{N}{2}]), the most clearly detected features in our spectra are the \ion{N}{2} lines. 
As \citet{2015MNRAS.454.4293P} note,
\thisSN\ appears to be the first SN~Ibn to show these features, though they are commonly
found in the spectra of hot stars.  In our DEIMOS spectrum we are able to identify nine \ion{N}{2} absorption lines
(see \S\ref{sec:hires}), and for the strongest of these we track their
evolution throughout our spectral series.  In our $-2.9$ and +0.1\,d spectra, 
the 5667\,\AA\ and 5676\,\AA\ features are apparent in emission
around $v \approx 0$\,\kms, but they transition into absorption at $v \approx 680$\,\kms\ after maximum light --- see Figure~\ref{fig:NaII}.
The same trend is apparent in the \ion{N}{2} features at 6482\,\AA\ and 7215\,\AA\ (Figure~\ref{fig:spec}).

We find a strong emission feature centred around 4650\,\AA\ in
our earliest spectrum.  In Figure~\ref{fig:spec} we label it as a blend of \ion{N}{3}, \ion{C}{3}, and \ion{He}{2}, a blend that
has been observed and modeled in the spectra of a few other young, interacting SNe \citep[e.g.,][]{2014Natur.509..471G,2015ApJ...806..213S,2015MNRAS.449.1921P}.
The 4650\,\AA\ emission feature fades rapidly.  Given the clear detection of \ion{N}{2} at all epochs we believe
the identification of \ion{N}{3} to be very robust, but whether the blue wing of that emission feature is caused by \ion{C}{3} or \ion{He}{2}
is more difficult to determine.  The \ion{He}{2}\,$\lambda$5411 line is extremely weak in \thisSN, which argues that 
\ion{C}{3} must be contributing to the 4650\,\AA\ feature, but discerning between these ions likely requires detailed modeling and
is beyond the scope of this paper.

\subsubsection{A Trace of Hydrogen?}
\label{sec:hydrogen?}

Inspection of our two-dimensional (2D) spectral images reveals host-galaxy emission features at H$\alpha$, H$\beta$, and H$\gamma$,
and there is no indication of either absorption or emission features intrinsic to \thisSN~for any of those lines.
However, our spectra at $-2.1$\,d and +3.1\,d do show an absorption feature near the expected wavelength of H$\delta$, 
decelerating over those 5 days from $-1060 \pm 75$\,\kms~(blueshifted) to $160 \pm 65$\,\kms~(redshifted relative to the rest frame of \thisSN,
consistent with the rest frame of NGC~2388) and then disappearing by +6.1\,d.
Though this feature is difficult to disentangle from the significant noise at these wavelengths,
it appears to show weak P-Cygni emission in our first observation.

NGC~2388 shows some H$\delta$ absorption, and the feature from +3.1\,d is probably associated with the galaxy (this spectrum
has undergone host-galaxy subtraction as described in \S\ref{sec:spec_data}).  However, in our $-2.1$\,d spectrum \thisSN~was
well-separated from the host and we performed no galaxy subtraction.
Along with the $\sim$1000\,\kms\ Doppler shift and faint P-Cygni feature, this indicates that the weak H$\delta$
feature in the earliest spectrum of \thisSN~arose within the same CSM as the \ion{He}{1} and other features described above.

Several SNe have been found to exist between the IIn and Ibn subclasses, exhibiting weak hydrogen features
\citep[e.g., SNe 2005la, 2010al, and 2011hw;][]{2008MNRAS.389..131P,2015MNRAS.449.1921P}.
Remarkably, SN~2010al had an H$\delta$ feature of comparable strength to its H$\alpha$ emission while showing only faint traces
of H$\beta$ and H$\gamma$ emission --- perhaps similar physics governs the hydrogen emission lines in \thisSN.
Regardless of this putative H$\delta$ feature, the amount of hydrogen in the CSM surrounding
\thisSN~must be vanishingly small.

\subsubsection{Higher Resolution}
\label{sec:hires}

\begin{figure*}
\includegraphics[width=\textwidth]{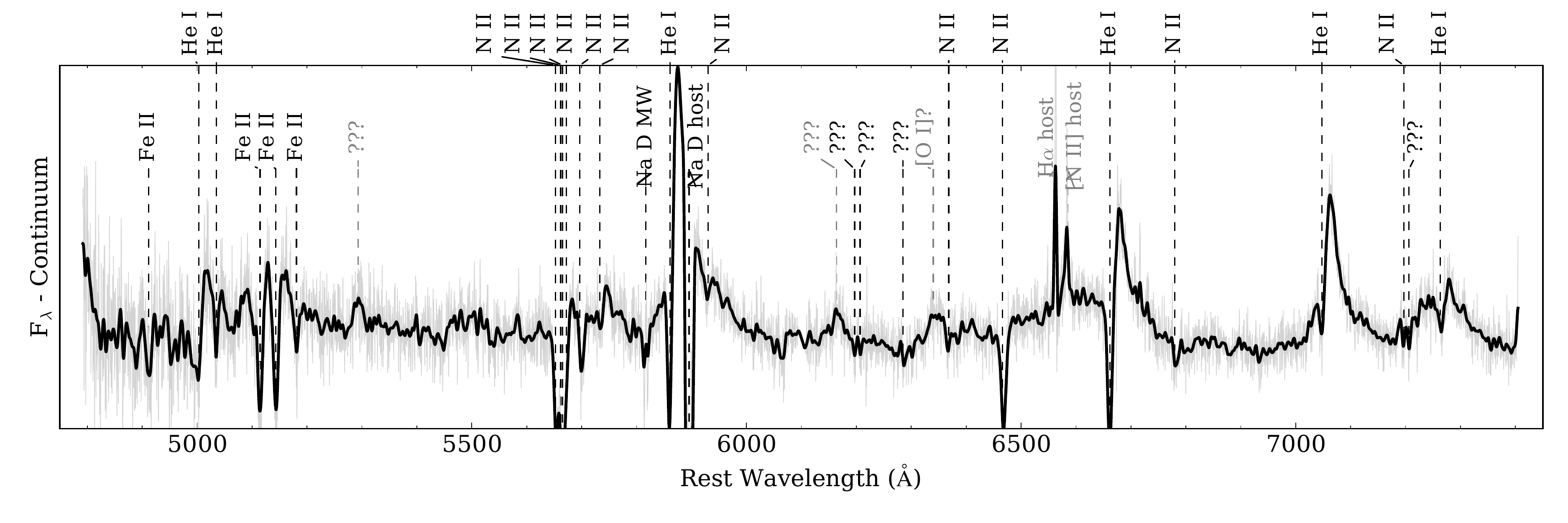}
 \caption{ The DEIMOS spectrum from Feb.\ 27 in detail.  A simple polynomial continuum fit has been removed.
    The full spectrum is shown in grey in the background while the black line displays the spectrum smoothed with a 10\,\AA\ Gaussian kernel.
    Identified emission lines are shown with grey dashed lines and text; absorption lines are in black.
    All absorptions are indicated with a 745\,\kms\ blueshift but emission lines are marked at rest velocity.
\label{fig:lineIDs}}
 \end{figure*}

Our DEIMOS spectrum from Feb. 27 ($R \approx 5000$) reveals
fully resolved absorption and emission lines from a host of ions in the spectra of \thisSN; see Figure~\ref{fig:lineIDs}
and Table~\ref{tab:lines}. We include measured positions and widths for several lines for which we have no good identification;
it is especially difficult to differentiate between \ion{Fe}{3} ($E_i =  30.651$\,eV) and \ion{Fe}{2} ($E_i = 16.20$\,eV).\footnote{All 
ionisation energies and line wavelengths were found through the NIST Atomic Spectra Database (ASD): \url{http://www.nist.gov/pml/data/asd.cfm}.}
In our spectrum we find at least four Fe absorption lines between 4900 and 5200\,\AA\ (in the SN rest frame).
The strongest two were identified as \ion{Fe}{3} by \citet{2015MNRAS.454.4293P}, but several known \ion{Fe}{2} lines occur at very similar wavelengths,
and we also identify weak features at 4912.5 and 5180.3\,\AA\ which are most plausibly interpreted as \ion{Fe}{2}\,$\lambda$4924 and \ion{Fe}{2}\,$\lambda$5195.
However, these identifications are tentative at best, given the nondetection of other expected \ion{Fe}{2} lines and the odd relative strengths of these
features, and so we caution the reader against overinterpreting the Doppler velocities presented in Table~\ref{tab:lines}.

\begin{table}
\caption{Absorption Lines in the DEIMOS Spectrum}
\begin{minipage}{\columnwidth}
\begin{center}
\label{tab:lines}
\begin{tabular}{ l | c c c c }
\hline
 Ion &  Rest $\lambda$  &  Observed $\lambda$ &  Offset  &  FWHM \\
  &   (\AA) &  (\AA) &   (\kms) &  (\kms)  \\
\hline
\ion{He}{1}  &  5015.7  &  5002.6  &  778.8  &  586.4  \\
\ion{He}{1}  &  5047.7  &  5034.6  &  783.2  &  267.8  \\
\ion{He}{1}  &  5875.6  &  5860.2  &  789.2  &  488.7  \\
\ion{He}{1}  &  6678.2  &  6661.7  &  737.2  &  470.1  \\
\ion{He}{1}  &  7065.2  &  7047.9  &  733.7  &  432.9  \\
\ion{He}{1}  &  7281.3  &  7265.4  &  655.9  &  404.0  \\
\ion{N}{2}  &  5666.6  &  5653.0  &  721.4  &  232.5  \\
\ion{N}{2}  &  5676.0  &  5665.4  &  559.2  &  433.1  \\
\ion{N}{2}  &  5710.8  &  5699.9  &  570.4  &  407.2  \\
\ion{N}{2}  &  5747.3  &  5734.8  &  651.6  &  367.5  \\
\ion{N}{2}  &  6482.1  &  6468.1  &  643.4  &  428.1  \\
\ion{N}{2}  &  5941.6  &  5927.5  &  711.5  &  600.7  \\
\ion{N}{2}  &  6796.6  &  6781.8  &  654.2  &  337.9  \\
\ion{N}{2}  &  7214.7  &  7195.9  &  783.5  &  173.7  \\
\ion{N}{2}  &  6384.3  &  6366.7  &  829.6  &  253.7  \\
\ion{Fe}{2}\footnote{\label{modref}It is difficult to determine exactly which \ion{Fe}{2} or
    \ion{Fe}{3} transition is responsible for these lines, and so the listed Doppler offsets are only tentative.}  &  4923.9  &  4912.5  &  698.0  &  677.2  \\
\ion{Fe}{2}\footref{modref}  &  5126.2  &  5114.9  &  658.1  &  514.5  \\
\ion{Fe}{2}\footref{modref}  &  5157.3  &  5143.2  &  819.4  &  584.9  \\
\ion{Fe}{2}\footref{modref}  &  5195.5  &  5180.3  &  873.2  &  545.7  \\
?  &  ---  &  6197.2  & --- &  166.9  \\
?  &  ---  &  6207.3  & --- &  152.7  \\
?  &  ---  &  6285.9  & --- &  153.8  \\
?  &  ---  &  7206.5  & --- &  190.5  \\
\hline
\end{tabular}
\end{center}
\par
Identifications and properties of the absorption lines identified
in our DEIMOS spectrum (after correcting for $z = 0.013161$).  All lines were fit by a Gaussian with
a linear approximation to the nearby continuum, and
rest wavelengths were downloaded from the NIST ASD. Characteristic error bars are $\sim0.6$\,\AA\ 
(30\,\kms) for the observed wavelengths and $\sim1.6$\,\AA\ for the Gaussian FWHM (85\,\kms), as estimated
by calculating the 95\% confidence intervals for the \ion{N}{2}\,$\lambda$5711 line using MCMC techniques. 
\end{minipage}
\end{table}

We also find three unidentified emission features in our DEIMOS spectrum that are not from host-galaxy contamination,
falling near 5290, 6165, and 6340\,\AA\ and exhibiting widths similar to those of the \ion{He}{1} lines.
None of these is clearly detected in our other spectra or the spectra published
by \citet[though generally they exhibit a S/N too low to rule them out]{2015MNRAS.454.4293P}.
The feature at 6340\,\AA\ is suggestive of the [\ion{O}{1}]\,$\lambda\lambda$6300, 6364 doublet that regularly arises in
nebular spectra of stripped-envelope SNe \citep[Types Ib/c; e.g.,][]{1997ARA&A..35..309F,2001PASP..113.1155M},
but this identification is only tentative.

\subsubsection{The Na\,D Lines}
\label{sec:csm}

 \begin{figure}
 \includegraphics[width=\columnwidth]{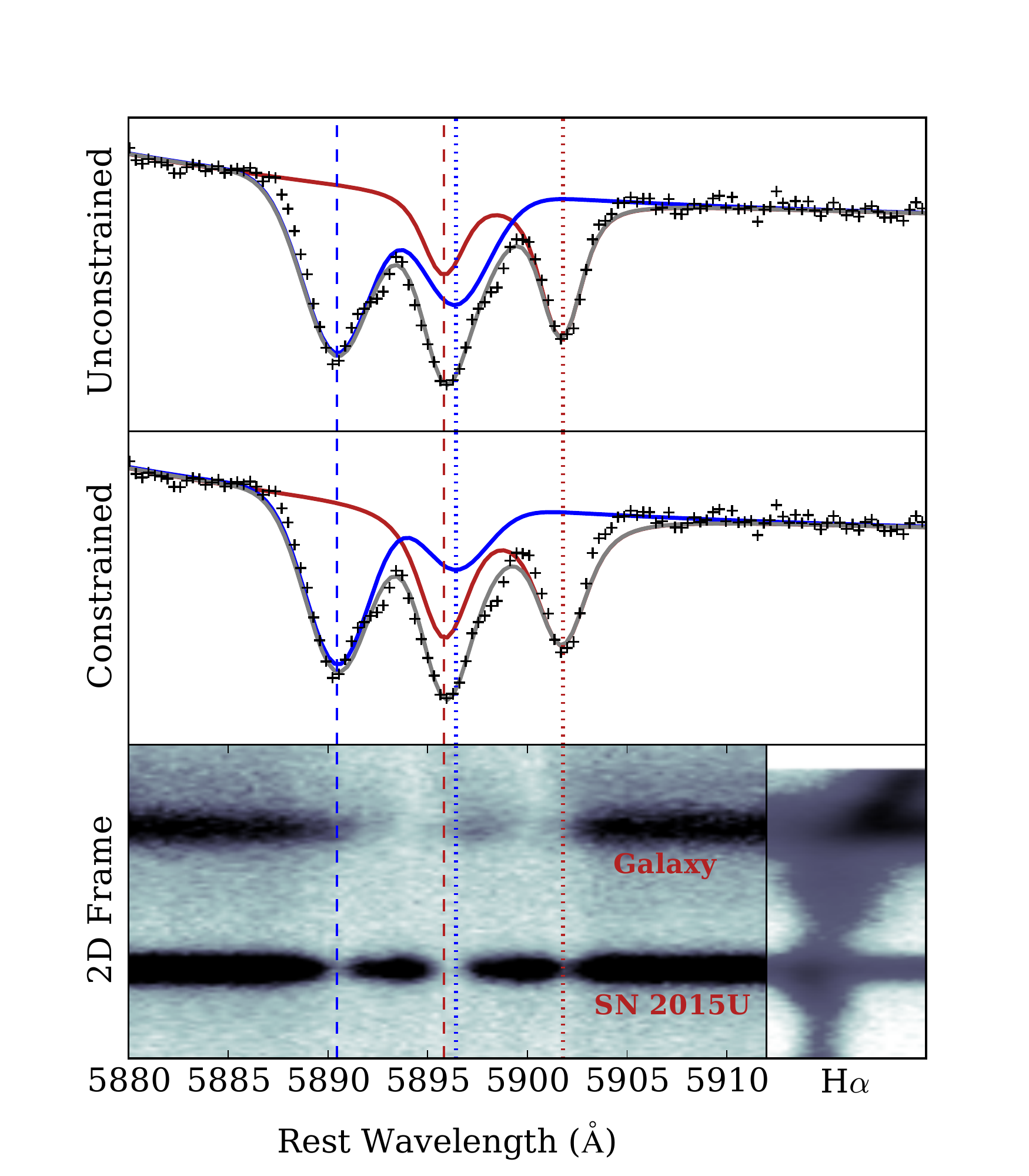}
 \caption{The Na\,D features in our DEIMOS spectrum of \thisSN\ and NGC~2388.  The top panel plots the observed spectrum of \thisSN\ 
    with our best-fit absorption profiles while the middle panel shows the result after introducing
   the additional constraint that D$_2$/D$_1$ $\geq 1.0$ (see \S\ref{sec:csm}).
    The shorter-wavelength doublet is shown in blue and the longer-wavelength one in red, with the central wavelengths indicated by
     dashed and dotted vertical lines for D$_2$ and D$_1$, respectively.  Their sum is given in grey.
    The bottom panel shows the 2D frame of our DEIMOS observation in this region, including both the light from the host and the SN,
    as well as a cutout of the H$\alpha$ emission line (plotted with the same Doppler $\delta v$ per pixel) to indicate the shape
    of the galaxy's rotation curve.
 \label{fig:nad} }
 \end{figure}

Our DEIMOS spectrum of \thisSN\ reveals a remarkable set of Na\,D absorption lines --- see Figure~\ref{fig:nad}.
In addition to the 
MW doublet we find several overlapping lines from NGC~2388 along our sight line, though as we show
below, there is no evidence that any of these lines originate within the CSM immediately surrounding \thisSN.
The astronomical Na\,D doublet has been studied in great detail (it was one of the original Fraunhofer lines)
and the relative strength of the doublet has long been used to measure the interstellar abundance
of neutral sodium: the Na\,D$_2$ line is generally observed to be stronger than the D$_1$ line by a factor ranging from 1.0 to
2.0, for high and low column densities of \ion{Na}{1}, respectively
\citep[e.g.,][]{1948ApJ...108..242S,1973ApJ...182..481N,1988Obs...108...44S}.
The MW Na\,D doublet in our spectra exhibits the expected behaviour, with an EW ratio of D$_2$/D$_1$ $\approx 1.2$,
but the host-galaxy lines do not.  

There are at least two distinct and overlapping Na\,D absorption doublets near the host-galaxy's rest
frame in the spectrum of \thisSN.
Figure~\ref{fig:nad} shows them (as well as model fits) in detail.  We model the absorption complex by overlaying
the modified Lorentzian emission-line profile (see \S\ref{sec:lineprofiles}) with two
doublets of Voigt absorption profiles (though we note that some of these lines may be unresolved and
saturated). D$_1$ and D$_2$ lines from a single doublet are forced
to have the same velocity properties and to be separated by 5.97\,\AA, but they are allowed independent strengths.

In the top panel of Figure~\ref{fig:nad} we show the result when we allow the relative strengths of the doublet (i.e.,\ D$_2$/D$_1$) to vary freely.
In the rest frame of the SN, the bluest doublet falls at $\lambda\lambda$5890.45, 5896.35 $\pm 0.066$\,\AA. 
This is a blueshift of only $\sim25$\,\kms\ from the SN rest frame, and so this
doublet is likely to arise from the ISM of NGC~2388 along our sight line to \thisSN.
The second (redder) doublet, however, falls at $\lambda\lambda$5895.80, 5901.69 $\pm$ 0.063\,\AA\ --- redshifted from the SN by almost 300\,\kms,
and even redshifted from the galaxy core by more than 100\,\kms.  This is within the velocity range of NGC~2388's
rotation curve, but \thisSN\ is located well away from the receding spiral arm and the bulk of the galaxy's receding material (see Figure~\ref{fig:finder}).
In addition, the second Na\,D doublet shows a strength ratio of D$_2$/D$_1$ $\approx 0.5$, 
well outside the commonly observed values of $\sim1.0$--2.0 (in contrast, the blue doublet exhibits a reasonable D$_2$/D$_1$ $\approx 1.5$).

If, instead, we constrain the doublet ratio to D$_2$/D$_1$ $\geq 1.0$, we can also obtain a reasonable fit, though this introduces noticeable
discrepancies in the reddest part of the feature --- see the middle panel of Figure~\ref{fig:nad}.
In addition, this forces the doublet ratio of the bluer doublet outside of the range of normal values: D$_2$/D$_1$ $\approx 2.5$.
Adding more components to our model fit will not solve this quandary (though the data do
show a shoulder on the bluer doublet, suggesting a third absorption component).
The implications for the \ion{Na}{1} gas (and the dust) in NGC~2388 are not clear.

Given the spectral signatures of dense CSM
surrounding \thisSN, it is natural to wonder whether some of the Na\,D absorption arises within the local CSM.
Variation in the Na\,D features over the timescale of the SN evolution would be a clear signature of local absorption \citep[e.g.,][]{2007Sci...317..924P};
however, as Figure~\ref{fig:HeI} shows, no such variation is apparent in our data.

The bottom panel of Figure~\ref{fig:nad} displays a cutout of the 2D DEIMOS spectrum around the Na\,D feature.
Both \thisSN\ and NGC~2388 are clearly visible, and we also show the rotation curve of the galaxy via a cutout of the H$\alpha$ emission line.
The Na\,D absorption of NGC~2388's own ISM is seen against the stellar light of the galaxy, and
this galactic self-absorption roughly covers the velocity range
of the two components observed along the line of sight toward \thisSN.
It appears that the anomalous red component described above obscures the host galaxy as well as the SN,
providing further evidence that this component is unlikely to arise within the CSM of \thisSN.

\subsubsection{Spectropolarimetry}
\label{sec:specpolAnalysis}

\begin{figure}
\includegraphics[width=\columnwidth]{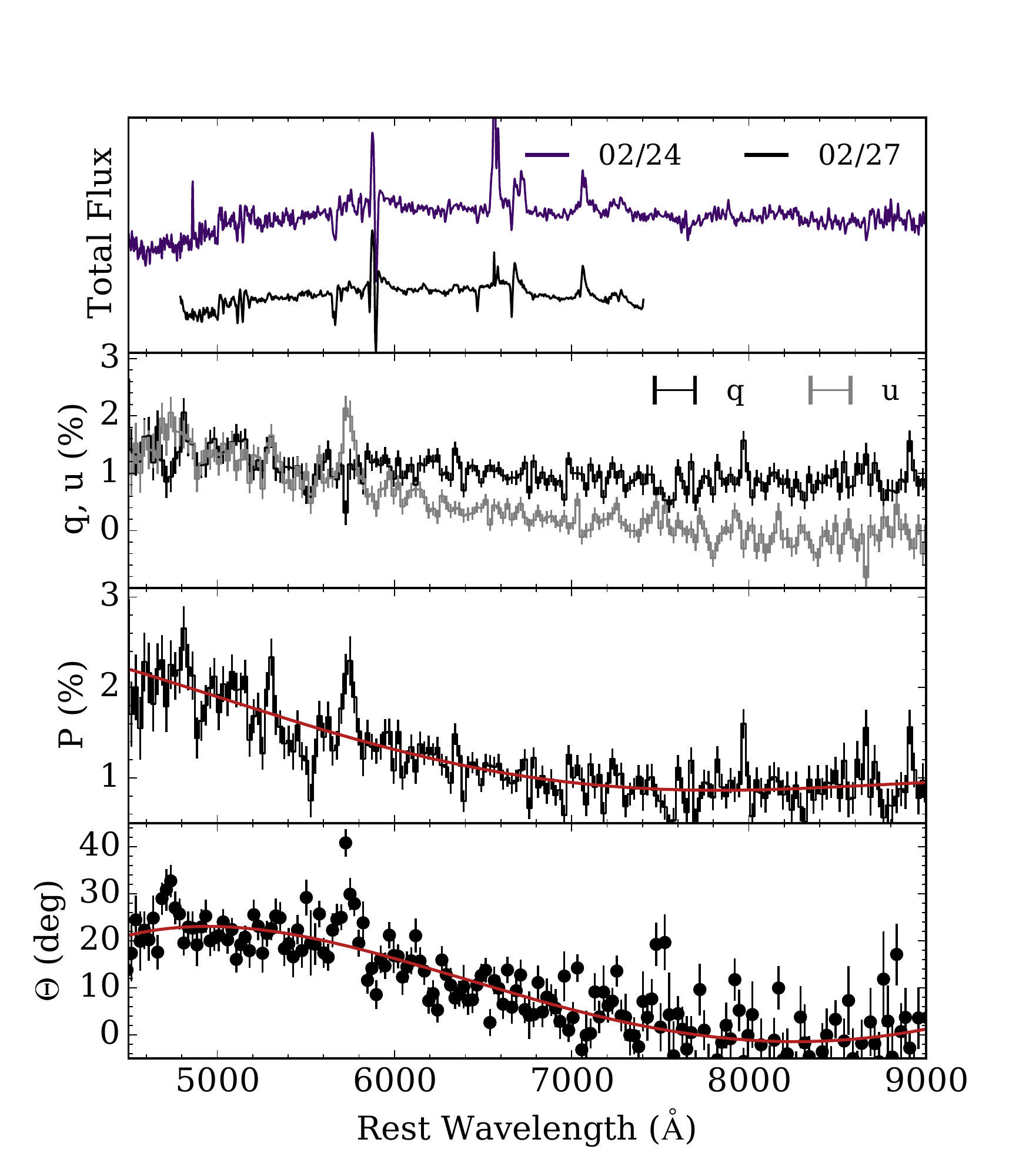}
\caption{Our spectropolarimetric observations of \thisSN.  From top to bottom:
    total-flux spectra near $V$-band peak and in higher resolution (for visual comparison),
    $q$ and $u$ Stokes parameters,
    $P$ with a fourth-order polynomial fit showing the overall trend,
    and $\theta$ with a similar polynomial fit (after discarding outliers).
   The spectropolarimetric data from three nights of observations have been coadded and binned to 25\,{\AA}.
  \label{fig:specpol}}
\end{figure}

We did not observe significant evolution in the polarization of \thisSN\ between our three epochs, so we coadded all of the data to increase 
the S/N; the results are illustrated in Figure~\ref{fig:specpol}. The polarization spectrum of \thisSN\ appears to be dominated 
by ISP associated with the host galaxy NGC~2388, showing a strong increase in continuum polarization toward shorter wavelengths with a value of 
$P \approx 2.5$\% at 4600\,{\AA} and decreasing to $P \approx 1.0$\% near 7000\,{\AA}. This behaviour is dissimilar to what is typically observed for the
MW ISP, which exhibits a peak and turnover near 5500\,{\AA}. Instead, the observed behaviour is reminiscent of the ISP produced by the host galaxies of 
several SNe~Ia, including SNe~1986G, 2006X, 2008fp, and 2014J \citep[][see their Figure 2]{2015A&A...577A..53P}. The continuous rise in $P$ beyond the
$B$ and $U$ photometric passbands has previously been interpreted as evidence for scattering by dust grains smaller than those characteristic of the MW disk ISM. 

Interestingly, we observe a significant wavelength dependence for the position angle ($\theta$), which has a value of $\sim25^{\circ}$ at 4600\,{\AA} 
and monotonically trends toward $\sim10^{\circ}$ near 7000\,{\AA}. In the MW, the ISP's position angle is generally flat. However, 
a wavelength dependence in $\theta$ has been observed along particular lines of sight toward star-forming regions at distances greater than 0.6\,kpc 
\citep[e.g.,][]{1965AJ.....70..579G,1966AJ.....71..355C}, and has been explained as the result of photons traversing multiple clouds or scattering media that 
exhibit various sizes for the scattering particles as well as various orientations for the interstellar magnetic field. 

A similar interpretation is plausible here: along the line of sight within NGC~2388 toward \thisSN\ there are likely to be multiple separate components
of dusty scattering media having different grain sizes and/or different magnetic field orientations. This possibility is particularly interesting considering
 the complex superposition of multiple \ion{Na}{1}\,D absorption doublets that we see in our high-resolution DEIMOS spectrum.
If the continuum flux spectrum of \thisSN\ is devoid of broad SN features because of high optical-depth CSM, then one might suspect there also to be a separate and
distinct scattering component associated with this CSM, if it is dusty. 
The subsequent re-scattering of this light as it traverses the dusty ISM of the host could provide a means for producing the observed wavelength dependence of $\theta$, 
if there is a difference in grain sizes between these multiple scattering media.
Although this scenario is physically plausible, the spectropolarimetric data cannot discriminate between a CSM$+$ISM scattering combination
and multiple components of host ISM, and our analysis of the
total-flux \ion{Na}{1}\,D features indicates that they are likely associated with ISM (see \S\ref{sec:csm}).

Finally, there is a line-like feature near 5820\,{\AA} which might naively be interpreted as a signature of intrinsic polarization associated with \thisSN\ and
with the \ion{He}{1} / \ion{Na}{1} transition. However, the MW \ion{Na}{1}\,D doublet falls at the same wavelength.
Although relatively weak compared to the redshifted host-galaxy Na\,D absorption, the results of our spectral arithmetic in the vicinity of this 
poorly resolved doublet profile might have created a spurious artifact in the final coadded dataset mimicking the shape of a polarized line feature. 
Indeed, in each individual epoch, the polarized spectra near this feature appear to suffer from systematic noise, so we are reluctant to attribute
this feature to the SN itself.

\subsection{Photometry}
\label{sec:phot_analysis}

\begin{figure}
 \includegraphics[width=\columnwidth]{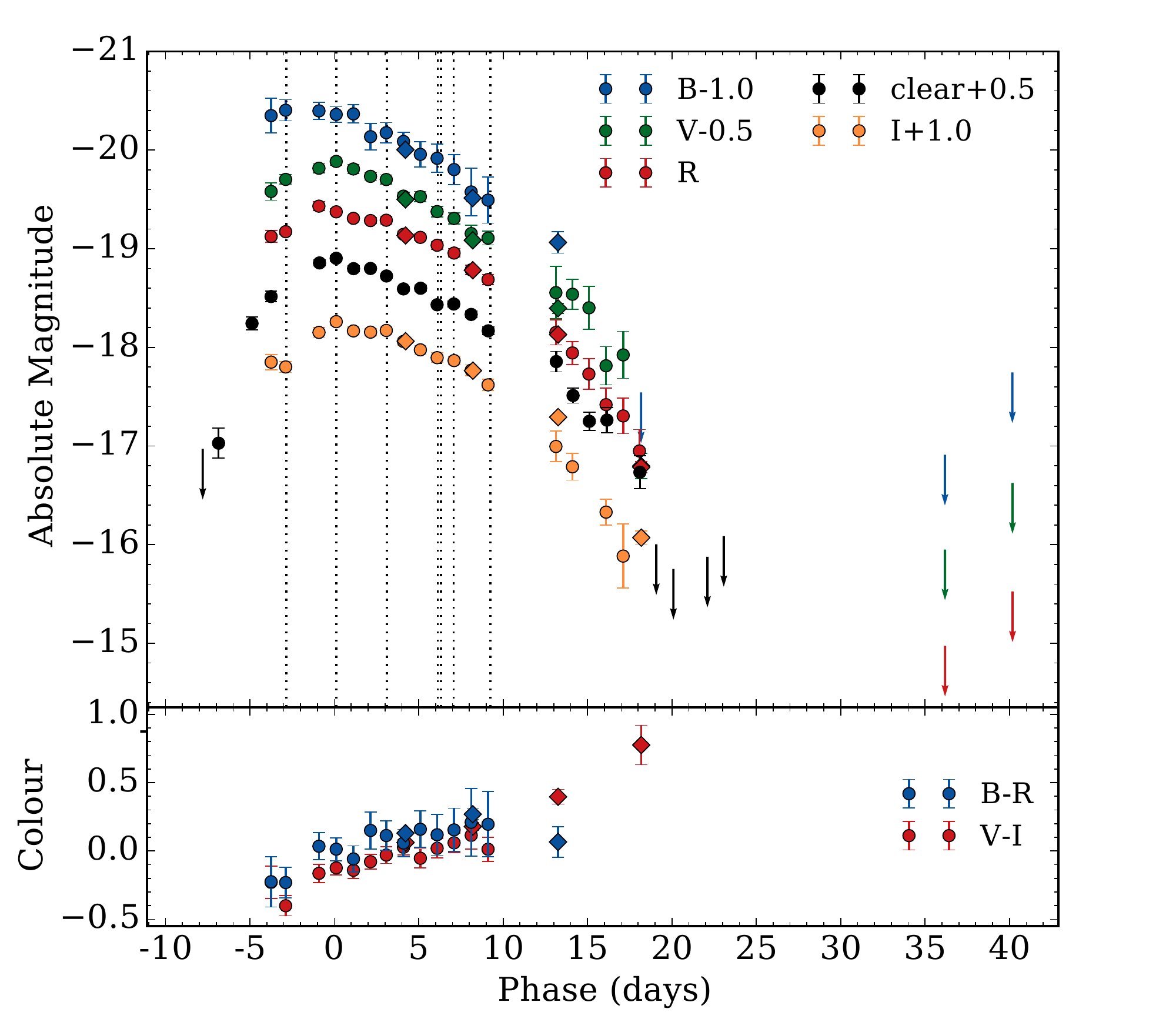}
 \caption{The KAIT and Nickel light curves (top) and colour curves (bottom) of \thisSN, after
    correcting for MW dust absorption and host absorption.
    Nickel data are shown with diamonds and KAIT data with circles, and
    the dates of our spectral observations are indicated in the top panel with dashed vertical lines.
    Host-galaxy contamination becomes significant in our KAIT photometry around +10\,d, and so
    we only show the Nickel data beyond that date in the lower panel (we include all data
    in the upper panel). We apply vertical offsets to every passband except $R$ in the top panel, to enable comparisons.
     \label{fig:phot_correct}}
 \end{figure}

The optical light curves of \thisSN\ show that it was a remarkably luminous and rapidly evolving event ---
Figure~\ref{fig:phot_correct} shows our photometry corrected for host-galaxy dust reddening.
With a peak absolute magnitude of $\lesssim -19$\,mag at optical wavelengths, 
a rise time of $\lesssim 10$\,d, a time above half-maximum of $t_{1/2} \approx 12$\,d, and a decline rate of nearly 0.2\,mag\,day$^{-1}$ after peak,
\thisSN\ was more than a magnitude brighter than most stripped-envelope SNe and evolved much more rapidly \citep[e.g.,][]{2011ApJ...741...97D,2014ApJS..213...19B}.

Though the first unfiltered ($clear$) KAIT detection of \thisSN\ was on Feb.\,11, and it went undetected by KAIT on Feb.\,10 ($> 18.4$\,mag),
\citet{2015MNRAS.454.4293P} present detections from Feb.\,9 and 10 at $R = 18.62 \pm 0.26$ and $18.14 \pm 0.30$\,mag, respectively.
Unfortunately, the location does not appear to have been observed in the days prior and there are not deep upper limits constraining the
explosion date further.  \thisSN\ rose quite rapidly, so we adopt a tentative explosion date one day before the
first detection published by \citet{2015MNRAS.454.4293P}:  $t_{\rm exp} \approx 57062$\,MJD (Feb.\,8).  This provides us with
a rise time for \thisSN\ of $t_{\rm rise} \approx 9$\,days.
\citet{2015IBVS.6140....1T} note that \thisSN\ is among the most rapidly evolving SNe known, with a decline
rate similar to those of SNe 2002bj, 2005ek, and 2010X.  We measure $\Delta M_{15} \approx 2.0$\,mag in the $R$ band,
but note that the decline rate increases ever more steeply after $\sim10$ days post-peak, and at all times the bluer passbands decline more rapidly than the red.
Simple linear fits indicate the following decline rates before and after $+10$\,d (in mag day$^{-1}$): 
$B_{\rm early} = 0.110 \pm 0.007$, $V_{\rm early} = 0.099 \pm 0.005$, $V_{\rm late} = 0.28 \pm 0.07$, 
$R_{\rm early} = 0.080 \pm 0.005$, $R_{\rm late} = 0.267 \pm 0.009$, $I_{\rm early} = 0.067 \pm 0.006$, and $I_{\rm late} = 0.26 \pm 0.04$
(uncertainties are statistical, and our data do not constrain the late $B$-band decline).

\thisSN\ is one of the nearest SNe~Ibn to date \citep{2016MNRAS.456..853P}, but it is still relatively distant for direct progenitor studies
(and is obscured by the dust in NGC~2388). Regardless, the {\it HST} nondetections presented in 
\S\ref{sec:phot} can be used to place interesting constraints on the SN's progenitor.
We compared these limits to the MIST stellar evolutionary tracks \citep{2016ApJ...823..102C} at solar metallicity generated in the
WFC3/infrared bandpasses
(negligible photometric differences exist between NICMOS/NIC2 and WFC3/IR for {\it F110W} and {\it F160W}).
Based on these tracks we can eliminate single-star progenitors with initial masses $M_{\rm ini} \gtrsim 
9\, {\rm M}_{\odot}$ and $\lesssim 40\, {\rm M}_{\odot}$.
That is, the progenitor would have been either a low-mass star near the core-collapse limit or a highly massive evolved star, possibly in a luminous blue variable (LBV)
or Wolf-Rayet phase at the time of explosion. We did not interpret our upper limits with respect to existing binary evolution models, and we caution
that these results are somewhat dependent upon the uncertain properties of NGC~2388's dust population.

The structure near \thisSN's position in the 1\,yr {\it HST} images (Figure~\ref{fig:finder}) may be due to clumpy star-forming regions, perhaps associated with \thisSN.
NGC~2388 is a strongly star-forming and massive galaxy --- \citet{2015A&A...577A..78P} calculate an ongoing star-formation rate
of $\sim 40^{+9}_{-22}$\,M$_{\odot}$\,yr$^{-1}$ and a total stellar mass of 10$^{11.0 \pm 0.1}$\,M$_{\odot}$ --- and though \thisSN\ 
is not obviously associated with the brightest star-forming regions of the galaxy, our spectra do show emission lines from nearby \ion{H}{2} regions.
The clump may also be an artifact of the intervening dust lanes in NGC~2388; unfortunately 
there is essentially no colour information available from our images as the host-galaxy background dominates.

\subsection{Comparisons with Other Supernovae}

 \begin{figure}
 \includegraphics[width=\columnwidth]{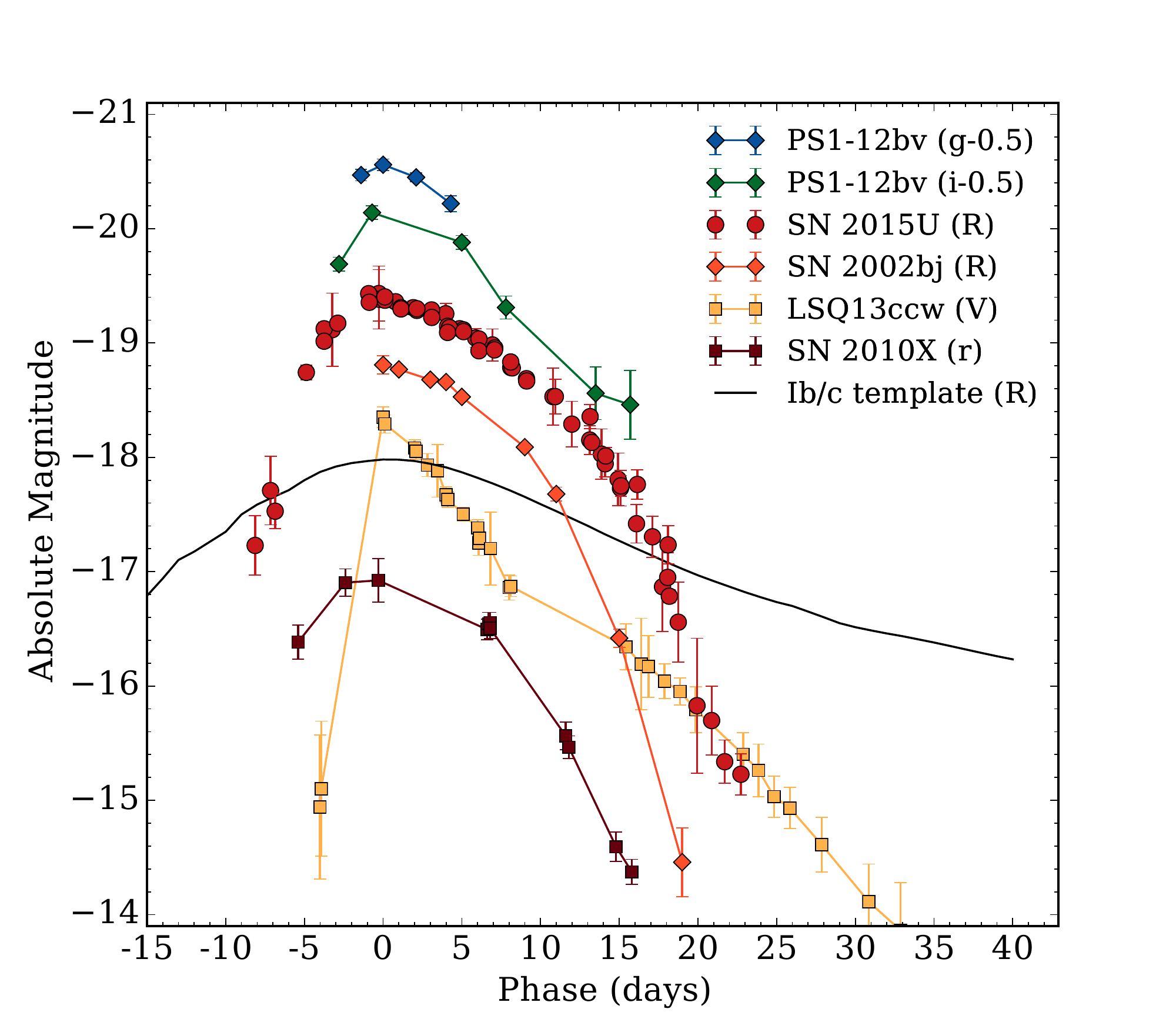}
 \caption{The extinction-corrected light curve of \thisSN\ compared to light curves of SNe~2002bj and 2010X \citep{2010Sci...327...58P,2010ApJ...723L..98K},
    the rapidly evolving SN~Ibn~LSQ13ccw \citep{2015MNRAS.449.1954P},
    and one of the rapidly evolving events from the PS1 sample \citep{2014ApJ...794...23D}. We also show in black the $R$-band SN~Ib/c template from
    \citet{2011ApJ...741...97D}.
    For \thisSN\ we include data from the {\it R} and {\it clear} passbands published here as well as the {\it r}-band photometry from \citet{2015MNRAS.454.4293P};
     a vertical offset of $-0.25$\,mag was applied to the {\it r}-band data to match the $R$ and $clear$ bands, enabling visual comparison.
     \label{fig:phot_comparisons}}
 \end{figure}

 \begin{figure*}
 \includegraphics[width=\textwidth]{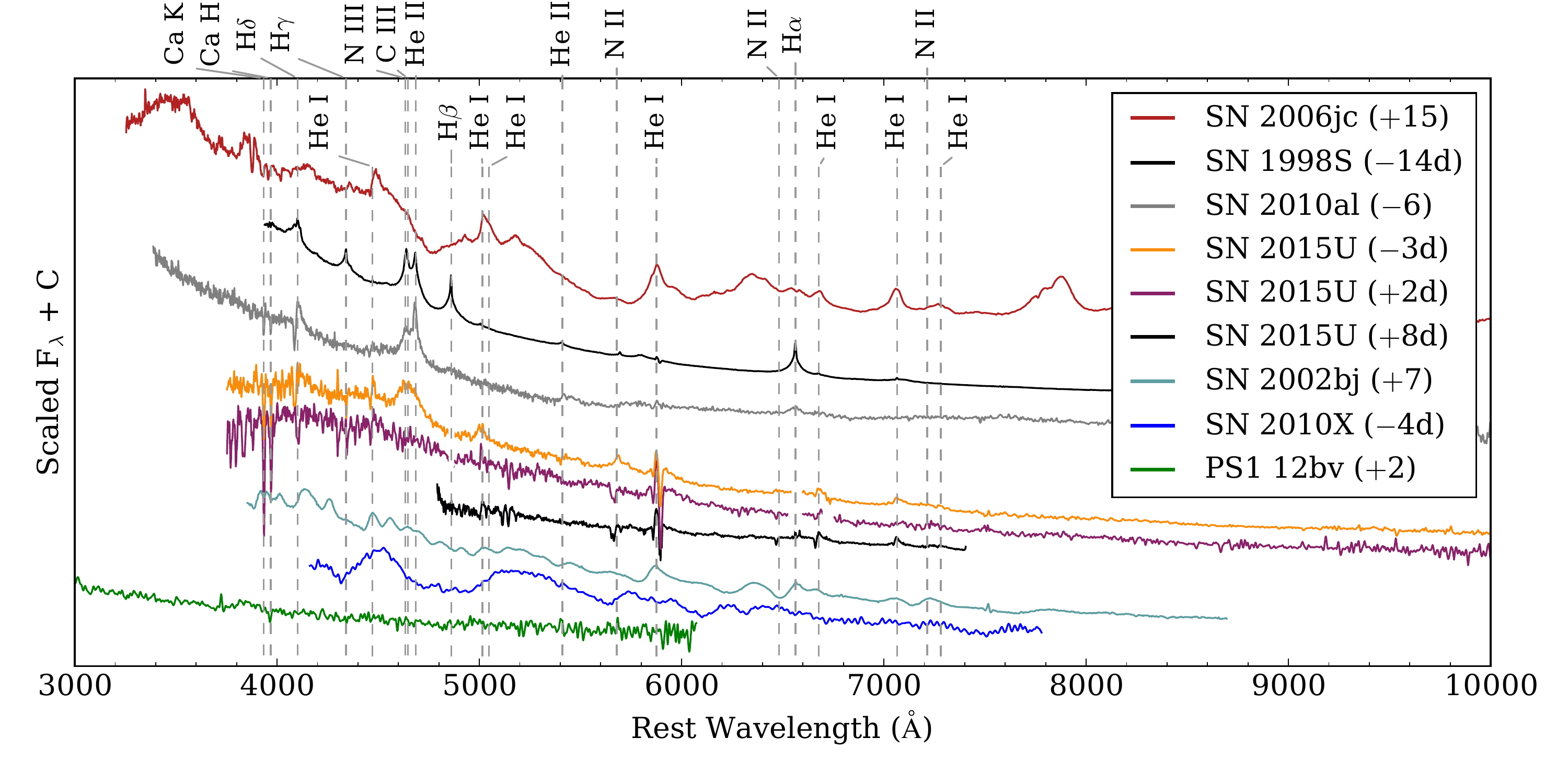}
 \caption{Spectra of \thisSN\ compared with spectra of the prototypical SN~Ibn~2006jc \citep{2007ApJ...657L.105F},
    the young SN~IIn~1998S \citep{2000ApJ...536..239L,2015ApJ...806..213S},
    a previously unpublished spectrum of the transitional SN Ibn~2010al \citep{2010CBET.2223....1S,2015MNRAS.449.1921P},
   the rapidly fading SNe~2002bj and 2010X \citep{2010Sci...327...58P,2010ApJ...723L..98K},
    and one of the rapidly fading events from the 
    PS1 sample \citep{2014ApJ...794...23D}.  Galactic emission features have been masked in the low-resolution spectra of
     \thisSN\ to facilitate comparisons.
     \label{fig:spec_comparisons}}
 \end{figure*}

In Figure~\ref{fig:phot_comparisons} we compare the light curves of \thisSN\ to the $R$-band light curves of SN~2002bj \citep{2010Sci...327...58P},
which was very similar though $\sim1$\,mag fainter, and SN~2010X \citep{2010ApJ...723L..98K}, which also
matches well but was $\sim2.5$\,mag fainter.  SNLS04D4ec and SNLS06D1hc, two of the rapidly rising transients discovered by the Supernova Legacy Survey (SNLS)
and presented by \citet{2016ApJ...819...35A}, also show similar light-curve behaviour but $\sim0.5$\,mag brighter than \thisSN\ (see their Figure~2).
In Figure~\ref{fig:spec_comparisons} we compare the spectra of these events: SN~2002bj was hydrogen deficient (SN~IIb-like, though not a good spectroscopic
match to normal SNe~IIb) with a strong blue continuum and P-Cygni features at higher velocities than observed in \thisSN, while SN~2010X showed no hydrogen
(SN~Ib/c-like, though again not a good spectroscopic match to normal SNe~Ib/c) with broad absorption features.
(Unfortunately, no spectra were obtained of the SNLS events.)

The light curves and spectra of \thisSN\ are quite similar to those of the rapidly evolving and luminous
transients discovered in the PS1 dataset and presented by \citet{2014ApJ...794...23D}.  Figure~\ref{fig:phot_correct}
includes the $g$ and $i$-band light curves of the relatively well-observed PS1-12bv from that sample;
it was discovered at $z = 0.405$, so the $g$ and $i$ passbands probe rest wavelengths around 3500\,\AA\ (about 100\,\AA\ shortward of $B$)
and 5350\,\AA\ (similar to $V$), respectively.  Without the $0.5$\,mag offset used in Figure~\ref{fig:phot_correct}, the $i$-band light curve
of PS1-12bv would overlie the $V$ band of \thisSN\ almost exactly, and the $g$-band curve would be somewhat brighter than the $B$ band
of \thisSN. There are no clear detections of any emission features in the spectrum of PS1-12bv, but it and the other
events from the PS1 sample were discovered at very large distances, and high-S/N
spectra do not exist for those events.  It is plausible that narrow emission lines of helium (and/or hydrogen) were present
but went undetected.

Figure~\ref{fig:spec_comparisons} shows that
\thisSN\ shares the blue colour and prominent \ion{He}{1} emission features of the canonical
SN~Ibn~2006jc \citep[e.g.,][]{2007ApJ...657L.105F,2008MNRAS.389..113P}.
However, SN~2006jc was discovered post-peak and evolved more slowly than \thisSN: the earliest extant spectrum of SN~2006jc
was taken after the event had faded $\sim1$\,mag from peak and the spectrum had already transitioned into a nearly nebular phase,
though a blue pseudocontinuum was apparent owing to ongoing interaction with the CSM \citep{2007ApJ...657L.105F}.
At later times, a second red continuum arose in SN~2006jc and the emission features became progressively more asymmetric and blueshifted,
evidence for dust formation with the SN system.  Increased absorption in the optical passbands by this dust, with re-emission in the
infrared, provided an explanation for the rapid increase in the optical decline rate observed for SN~2006jc after $\sim$50 days
\citep{2008MNRAS.389..141M,2008ApJ...680..568S,2008MNRAS.389..113P}.

In contrast, \thisSN\ presented a spectrum dominated by a single blue continuum through our last epoch of spectroscopy,
also taken after the event had faded $\sim1$\,mag from peak.  The $B-R$ and $V-I$ colour curves of \thisSN\ smoothly 
trend redward from a few days before (optical) maximum brightness to $\sim$2\,mag out onto the rapidly fading tail 
(at which point the SN fades below our detection threshold) --- see Figure~\ref{fig:phot_correct}.
The accelerating decline rate observed in the \thisSN\ light curves does not appear to be due to dust formation: one
would expect a redward knee in the colour curves if it were.  On the contrary, there is some evidence that the $B-R$ curve
of \thisSN\ was beginning to flatten in our last few epochs, though note that $B$-band data become sparse.

We also show the light curve of the peculiar and rapidly fading SN Ibn LSQ13ccw in Figure~\ref{fig:phot_comparisons} \citep{2015MNRAS.449.1954P}.
LSQ13ccw rose to peak extremely quickly ($t_{\rm rise} \approx 5$\,d) and then faded rapidly for $\sim$10 days, similar to \thisSN's behaviour.
However, it thereafter slowed in its decline, and the spectra of LSQ13ccw (not shown) exhibited both broad and narrow features (as do many, but not all, SNe~Ibn).
The knee in the light curve of LSQ13ccw is plausibly interpreted as evidence for ongoing energy
injection from ejecta/CSM interaction in the system --- no such knee is observed in the light curve of \thisSN, which instead appears to be consistent
with a single shock-breakout/diffusion event.

The early-time spectrum of \thisSN\ is similar to the very early spectra of the Type IIn SN~1998S
and the transitional Type IIn/Ibn SN~2010al, excepting the absence of hydrogen (though see \S\ref{sec:hydrogen?}
for a discussion of the tentative H$\delta$ feature in \thisSN).
All three events show a smooth, blue, and (approximately) blackbody continuum. 
The implied CSM velocity of \thisSN\ is higher than in SN 1998S and SN 2010al,
and \thisSN\ shows strong \ion{He}{1} lines while the other two events exhibit stronger \ion{He}{2}.
The \ion{N}{3}/\ion{C}{3} complex near 4500\,\AA\ is clearly detected in all three events ---
this feature dominates at very early times and disappears by peak (though that
evolution took place over different timescales for these SNe).

Despite the spectral similarities, these three SNe displayed diverse light-curve behaviour.
SN~1998S had a SN~IIL-like light curve with a long-lasting tail \citep[e.g.,][]{2000MNRAS.318.1093F},
while the evolution of SN~2010al was more similar to that of \thisSN\ though less rapid
\citep{2015MNRAS.449.1921P,2016MNRAS.456..853P}.  The $B-R$ colour curve of
\thisSN\ (as shown in Figure~\ref{fig:phot_comparisons}) is also notably similar to that of SN~2010al,
with a slow trend toward redder colours for most of the SN's evolution and then either leveling off
or perhaps even becoming more blue again beyond +10\,d and +20\,d for SNe 2015U and 2010al, respectively
\citep[][see their Figure 4]{2015MNRAS.449.1921P}.

The $\sim20$-day timescale for \thisSN's evolution and the increasing decline rate are reminiscent of the 
light curves expected for low-$^{56}$Ni explosions in helium-dominated or oxygen-dominated envelopes ---
see, for example, the models of \citet{2011MNRAS.414.2985D} and \citet{2014MNRAS.438..318K}.  In these models, 
the recombination of helium or oxygen produces a dramatic and rapid drop in opacity within the ejecta and an inward-moving recombination front.
\thisSN\ was much more luminous than the objects in these models, but perhaps it was an analogous event wherein
a recombination wave in the extended CSM produced a rapid fade from maximum luminosity as the
CSM cooled after shock breakout.
Similar situations have been observed in hydrogen-rich SNe:
there exists a subclass of SNe~IIn that shows light curves with a plateau likely produced via hydrogen recombination
within their extended CSM \citep[SNe 1994W, 2009kn, and 2011ht; e.g.,][]{2004MNRAS.352.1213C,2012MNRAS.424..855K,2013MNRAS.431.2599M}.

The narrow P-Cygni profiles and strong continuua of \thisSN's spectra are clear indications of dense CSM.
The diversity of ways in which such CSM can affect the light curves of SNe has been explored in detail by several authors
motivated by observations of SNe~IIn/Ibn, superluminous SNe, and rapidly fading SNe 
\citep[e.g.,][]{2010ApJ...724.1396O,2011ApJ...729L...6C,2014MNRAS.438..318K},
and comparisons of \thisSN's light curve to these models implies that \thisSN\ was a shock-breakout
event.  The high luminosity indicates that shock breakout occurred at a large radius and that a significant
fraction of the SN's kinetic energy was converted into light, while the lack of a long-lasting
light-curve tail shows that the CSM surrounding \thisSN\ was more shell-like than wind-like 
(there is no interaction-powered tail) and that relatively little $^{56}$Ni was produced
(there is no radioactively powered tail).

The rapidly fading light curves of SNe~2002bj and 2010X are remarkably similar to those of \thisSN, and we argue
that their spectral differences do not exclude the possibility that these three events were fundamentally quite similar.
By varying the radius of a putative opaque CSM shell in a simple model,
we can understand events spanning a range of luminosities and timespans for which the opaque CSM
reprocesses the SN flux.
Under the assumption that their light curves are all shock cooling curves without significant contribution from radioactive nickel,
the peak luminosities and timescales of these three SNe should be governed roughly by  
\begin{equation*}
t_{\rm SN} \propto E^{-1/6} M^{1/2} R^{1/6} \kappa^{1/6} T^{-2/3}\, ,
\end{equation*}
\begin{equation*}
L_{\rm SN} \propto E^{5/6} M^{-1/2} R^{2/3} \kappa^{-1/3} T^{4/3}\, ,
\end{equation*}
following \citet{2014MNRAS.438..318K},
adapted from \citet{2009ApJ...703.2205K} for hydrogen-free SNe and based on the 
analytic framework of \citet{1993ApJ...414..712P}.  $E$ is the energy of the SN explosion,
$M$ is the mass ejected, and $R$ is the effective radius (in this case, the radius reached by a significant amount of mass ejected prior to the final explosion).

In particular, consider the peak-luminosity equation, assuming these explosions are similar in all their properties except for the effective
pre-SN radius $R$, which would be determined by the time before explosion
and the speed at which any pre-SN material was ejected.
Inverting the equation, the pre-SN effective radius $R \propto L_{\rm SN}^{3/2}$. If we assume that the ejecta velocities of all three objects are similar,
this means that the ejecta from the SN itself will pass through the surrounding CSM in a time $t_{\rm interact} \propto R$,
and the length of time after explosion we expect to see narrow lines from this interaction is proportional to $R$. Comparing the
relative peak luminosities, we find that, since \thisSN\ exhibited narrow lines at least up to $\sim16.5$ days after explosion, SN\,2002bj should
have shown narrow lines at least until $\sim 5.5$ days after explosion and the even dimmer SN\,2010X would have shown narrow lines
at least until $\sim 0.52$ days. These times are well before the first spectra were taken of either SN 2002bj or SN 2010X. Typically,
we expect that to discover narrow lines in these rapidly fading SNe, we will either need to look at the brightest among them or catch them very early. The
luminosity (and timescale) may also depend on the explosion energies, ejected masses, and other properties of the SN, but these must be
disentangled with more sophisticated numerical approaches.  

\subsection{Temperature and Luminosity Evolution}
\label{sec:temp}

The extreme and uncertain degree of host-galaxy dust reddening toward \thisSN\ makes estimating bolometric
properties quite difficult.  In \S\ref{sec:dust} we assume that the emission from \thisSN\ is roughly blackbody 
in spectral shape; here we further discuss this assumption. 
It has long been known that the continua of young SNe~II exhibit ``diluted blackbody'' spectral energy distributions (SEDs), which (at optical wavelengths)
are similar to the Planck function at a lower temperature \citep[e.g.,][]{1986ApJ...301..220H,1996ApJ...466..911E}.
Though \thisSN\ is certainly not a SN~II, it is continuum-dominated, and in the sections above we present evidence
that \thisSN\ was shrouded in an optically thick CSM.
In addition, 
\citet{2011MNRAS.415..199M} show that modeled shock breakouts from the hydrogen-dominated CSM around red supergiants exhibit
roughly blackbody SEDs.
\citet{2011ApJ...728...63R} explore shock breakouts from He or He/CO stellar envelopes in detail; for He-dominated stellar
envelopes, they show that the colour temperature of the system deviates from the photosphere's temperature by a (time-dependent) factor 
of only $\sim20$\%, largely owing to diffusion effects as a helium recombination wave moves through the material.

Given the above discussion, we assume that \thisSN's emission was roughly blackbody and
we estimate the bolometric properties by fitting a dust-reddened Planck spectrum to our multiband photometry.
Using our almost nightly observations
of \thisSN\ in the {\it BVRI} passbands, we assembled the observed optical SEDs from $-5$\,d to +20\,d and 
find the best-fit blackbody temperature, radius, and luminosity using MCMC maximum-likelihood methods.
As described in \S\ref{sec:dust}, there are large uncertainties in the properties of the host galaxy's dust, and
these uncertainties produce similarly large uncertainties in the absolute value of the temperature at any given time.
In addition, our optical photometry largely probes the Rayleigh-Jeans tail of \thisSN's SED, so the implied 
bolometric corrections are large and uncertain.  Our MCMC-produced error bars include
these effects, and our photometric data do indicate significant temperature evolution (assuming $E(B-V)$ and $R_V$ do not change).
Because our data constrain the evolution of \thisSN\ more
strongly than they constrain the absolute values of any given parameter, it is useful to consider the implications for a
few assumed values of $E(B-V)$.

\begin{figure*}
 \includegraphics[width=\textwidth]{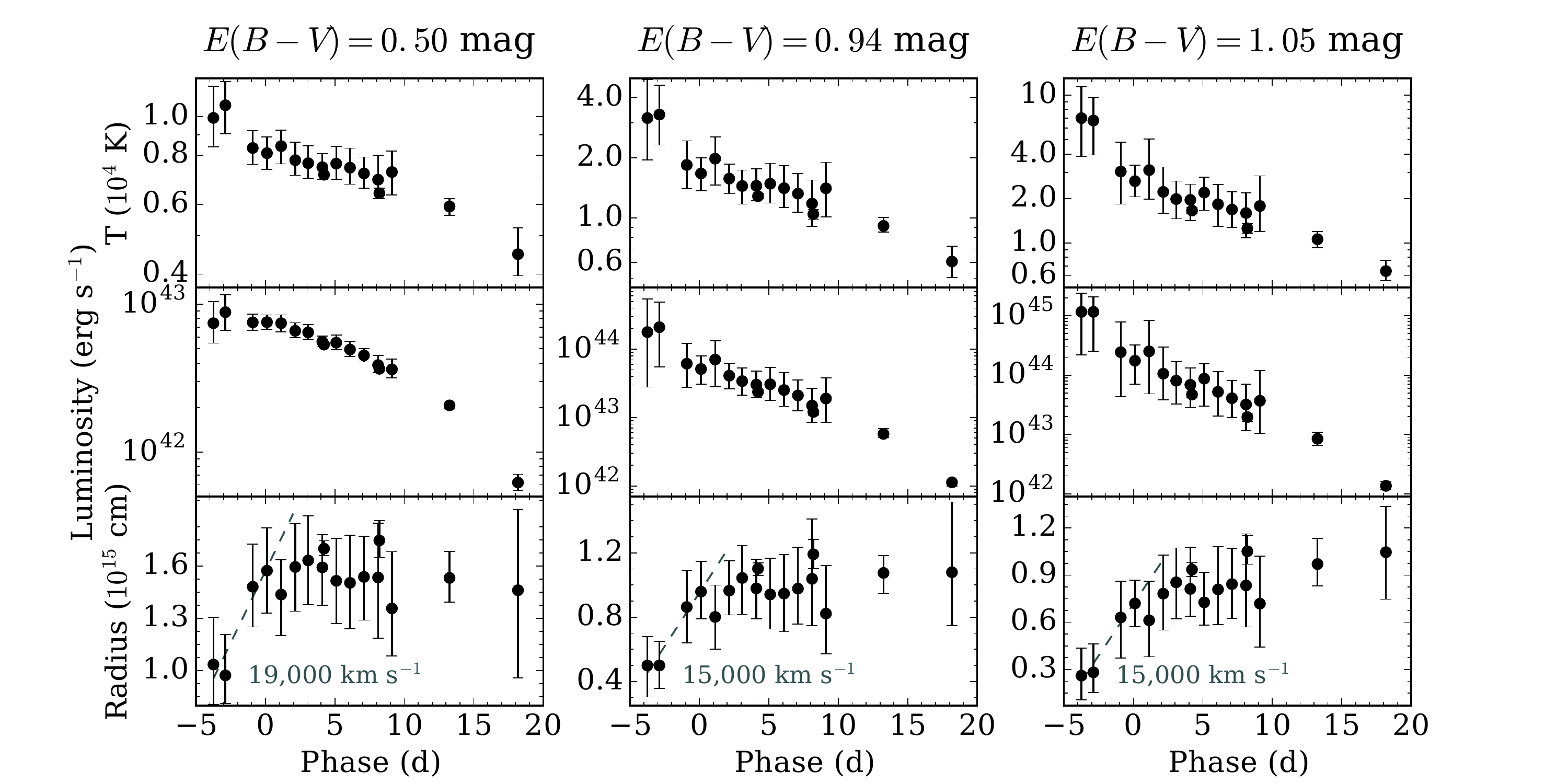}
 \caption{The best-fit blackbody temperature, bolometric luminosity, and radius evolution of \thisSN\ for three assumed values of $E(B-V)$
    as a function of phase relative to $V$-band peak.
    The leftmost column shows the evolution assuming a value of 0.5\,mag, the middle column assumes
    our preferred value of 0.94\,mag, and the rightmost column corresponds to 
a value of 1.05\,mag.  In the lower panels, we show linear fits to the radius evolution up to 
    $V$-band peak with a grey dashed line.
     All parameters are fitted using maximum-likelihood MCMC methods and error bars represent 95\% confidence intervals (but do not include
    errors caused by uncertainty in the degree of host-galaxy dust reddening).
     \label{fig:TLRt}}
 \end{figure*}

Figure~\ref{fig:TLRt} shows the blackbody temperature, bolometric luminosity, and radius evolution of \thisSN\ for
three different assumed values of $E(B-V)$, adopting $R_V = 2.1$.  In all cases, the best-fit temperature decreases over time, with the temperature
falling rapidly before (optical) peak and then decreasing slowly thereafter.  
The effective blackbody radius increases with a photospheric velocity of $v_{\rm phot} \approx 15,000$\,\kms\ until peak,
leveling off thereafter. Though this value of $v_{\rm phot}$ is similar to the characteristic velocities of ejecta in stripped-envelope SNe, our spectra 
do not show any material moving that quickly and, given our interpretation of \thisSN\ as a cooling shock-breakout event, we expect the photospheric radius to 
be approximately constant throughout these observations.  
In addition, Figure~\ref{fig:TLRt} may surprise the reader by indicating that the peak bolometric luminosity occurred several days before
the optical peak.  Though the bolometric corrections are at their largest and most uncertain at early times,
and we unfortunately do not know of any observations constraining the ultraviolet emission of \thisSN,
similar behaviour was observed from SN~Ibn~2010al with ultraviolet to near-infrared wavelength coverage \citep{2015MNRAS.449.1921P}.

In addition, we can constrain the temperature evolution of \thisSN\ independently of the continuum shape
through the relative line strengths of detected emission lines in our spectra.  Though detailed modeling is
beyond the scope of this paper, the pre-maximum detection and rapid fading of the \ion{N}{3}/\ion{C}{3}/\ion{He}{2} 4500\,\AA\ complex in our spectra of
\thisSN\ brings to mind the spectral evolution of SN~1998S (see Figure~\ref{fig:spec_comparisons}).  Detailed CMFGEN
models of SN~1998S while the 4500\,\AA\ feature was strong were
presented by \citet{2015ApJ...806..213S}, indicating a temperature
of $\sim$30,000\,K throughout most of the line-forming region.  The agreement between this value and the pre-maximum
temperatures found for \thisSN\ assuming $E(B-V) = 0.94$\,mag are heartening.

In Table~\ref{tab:intbolo} we present the integrated energy released and various values at $V$-band peak for
three assumed values of $E(B-V)$. Our uncertainty about the host galaxy's dust properties is the dominant source of error in this analysis,
and the spread over these three values of $E(B-V)$ indicates the range.  All values are estimated through simple
polynomial fits to the curves shown in Figure~\ref{fig:TLRt}.  To calculate $E_{\rm rad}$, we integrated from
day $-5$ to day +20 (the timespan of our almost nightly multiband photometric coverage), and for the other parameters
we list the value at $V$-band peak. Note that the bolometric luminosity peaks before the optical
maximum; again, see Figure~\ref{fig:TLRt}.

\begin{table}
\begin{minipage}{\columnwidth}
\begin{center}
\caption{Bolometric Properties of \thisSN}
\label{tab:intbolo}
\begin{tabular}{ l | c c c c }
\hline
 $E(B-V)$  &  $L_{\rm peak}$   &  $E_{\rm rad}$  &   $T_{\rm peak}$  &  $R_{\rm peak}$  \\
 (mag)         &  (erg\,s$^{-1}$)    &  (erg)                 &  (K)                   &  (cm) \\
\hline
0.5  & $7.3 \times 10^{42}$ & $9.3 \times 10^{48}$ & 8400 &  $1.5 \times 10^{15}$  \\
0.94 & $5.9 \times 10^{43}$ & $7.1 \times 10^{49}$ & 19,000 & $8.7 \times 10^{14}$ \\
1.05 & $2.3 \times 10^{44}$ & $2.9 \times 10^{50}$ & 30,000 & $6.7 \times 10^{14}$ \\
\hline
\end{tabular}
\end{center}
\par
Bolometric luminosity, radiated energy, blackbody temperature, and blackbody radius for the three assumed values of $E(B-V)_{\rm host}$
shown in Figure~\ref{fig:TLRt}.
\end{minipage}
\end{table}

\subsection{Nickel Content}

A slowly declining light-curve tail, powered either via CSM interaction or $^{56}$Ni decay, has been detected in some SNe~Ibn,
but many members of this subclass (including \thisSN) do not show one \citep[e.g.,][]{2007ApJ...657L.105F,2014MNRAS.443..671G,2016MNRAS.456..853P}.
For most SNe~I, Arnett's law can be used to estimate the total $^{56}$Ni synthesized by the explosion \citep{1982ApJ...253..785A}
based upon observed properties near the time of peak luminosity; however, that approach is not applicable for shock-breakout events.
If the opacity of the ejecta is well understood, the luminosity at late times can be used instead.
For example, \citet{2003ApJ...582..905H} analyzed Type IIP supernovae and,
given a luminosity $L_t$ at time $t$ and an explosion time $t_0$, estimated the $^{56}$Ni mass to be
\begin{multline*}
M_{\rm Ni} = (7.866 \times 10^{44})\, L_t \, \mathrm{exp} \left( \frac{(t - t_0)/(1+z) - 6.1}{111.26\,{\rm d}} \right)\, {\rm M}_{\odot}.
\end{multline*}
This equation assumes that $\gamma$-rays produced via radioactive decay are fully trapped and thermalized
by the ejecta, an assumption which (though reasonable for SNe IIP) has been shown not to be true for
most stripped-envelope SNe \citep[which fade faster than expected for complete trapping; e.g.,][]{2015MNRAS.450.1295W}.

The extreme CSM surrounding \thisSN\ further complicates the issue, as it may be contributing luminosity
via ongoing interaction at late times and perhaps even trapping a higher fraction of the $\gamma$-rays than is normal for CSM-free SNe Ib/c.
Though a robust measure of the amount of nickel created by \thisSN\ would require more sophisticated treatment,
we place a rough upper limit on the value by assuming that the ejecta+CSM system of \thisSN\ completely traps any
radioactively produced $\gamma$-rays and that there is no luminosity contribution from ongoing interaction beyond
our last detection at $\sim 20$\,days.

Our most constraining observation of a putative radioactively powered tail is a nondetection in the $R$ band at +36\,d.  
Our last multiband photometric measurement of the temperature yields $T = 6000 \pm 1000$\,K at +18\,d (see \S\ref{sec:temp}).
We adopt that temperature to calculate the blackbody luminosity for a source at our nondetection threshold,
but we note that the temperature is still changing at this time, possibly affecting our results.
This calculation yields an upper limit of $M_{\rm Ni} \lesssim 0.02$\,M$_{\odot}$ ---
quite low for SNe~I
\citep[$M_{\rm Ni,Ia} \gtrsim 0.4$\,M$_{\odot}$, $M_{\rm Ni,Ib/c} \approx 0.2$\,M$_{\odot}$;][]{2000A&A...359..876C,2011ApJ...741...97D},
but within the range of values observed for SNe~II
\citep[$M_{\rm Ni,II} \approx 0.0016$--0.26\,M$_{\odot}$;][]{2003ApJ...582..905H}.

\subsection{Progenitor Mass-Loss Rate}

We adopt the simple shock-wind interaction model of \citet{2011ApJ...729L...6C} to estimate the properties of 
the CSM surrounding \thisSN.  We assume that the CSM is spherically symmetric and follows a wind-like
density profile ($\rho \propto r^{-2}$), and we use the opacity of helium-dominated material in our calculations
($\kappa = 0.2$\,cm$^2$\,g$^{-1}$).   
\citet{2014ApJ...780...21M} solve the \citet{2011ApJ...729L...6C} model 
in terms of three observables: break-out radius $R_{\rm bo}$ (the radius at which radiation diffuses forward ahead of the shock), total radiated energy $E_{\rm rad}$, and light-curve rise time $t_{\rm rise}$.
We use $t_{\rm rise} = 9$\,days and, assuming $E(B-V)_{\rm host} = 0.94$\,mag, we adopt $E_{\rm rad} = 7.1 \times 10^{49}$\,erg from Table~\ref{tab:intbolo}.
For $R_{\rm bo}$ we take the first measured blackbody
radius from Figure~\ref{fig:TLRt}: $R_{\rm bo} = 5 \times 10^{14}$\,cm.  
The resultant mass-loss estimate is very large, $\dot M \approx 1.2$\,M$_{\odot}$\,yr$^{-1}$. 
However, in this model $\dot M \propto R_{\rm bo}^{-3}$, and there is reason to believe that our measurement of the blackbody
radius before peak brightness does not reflect the true $R_{\rm bo}$ (see \S\ref{sec:temp}).  If we instead use $R_{\rm bo} = 9 \times 10^{14}$\,cm,
the blackbody radius at $V$-band peak, we calculate $\dot M \approx 0.2$\,M$_{\odot}$\,yr$^{-1}$. 

Both of these values are several orders of magnitude above the most extreme mass-loss rates produced by steady winds from stars,
but they are not far from the time-averaged eruptive mass-loss rates from LBVs
\citep[which may well exhibit much higher instantaneous mass-loss rates;][]{2014ARA&A..52..487S}.
Observations of iPTF13beo, a recent SN~Ibn discovered by the intermediate Palomar Transient Factory (iPTF), implied an even higher (but short-lived)
mass-loss rate of $\dot M \approx 2.4$\,M$_{\odot}$\,yr$^{-1}$ immediately before the progenitor underwent core collapse
\citep[estimated via similar methods;][]{2014MNRAS.443..671G}.

If the CSM around \thisSN\ was launched from the surface explosively rather than through a steady wind, the assumption that
$\rho_{\rm csm} \propto r^{-2}$ is suspect.  In fact, the lack of an interaction-powered tail in the light curve of \thisSN\ 
indicates that the density profile cannot be wind-like: \citet{2011ApJ...729L...6C} show that ongoing interaction with
a wind-like CSM powers a tail with $L \propto t^{-0.6}$, and \thisSN\ fades more rapidly than that throughout its post-peak evolution.
The lack of any high-velocity features in our spectra argues that this outer CSM must be optically thick at least out to the radius
of the shock at the time of our last spectrum. Assuming $v_{\rm shock} \approx 20,000$\,\kms, the velocity of the 
fastest material in unshrouded stripped-envelope SNe, this produces an estimate for the CSM extent of $R \gtrsim 3 \times 10^{15}$\,cm.
Adopting the average \ion{He}{1} velocity measured from our spectra (745\,\kms), material at that radius was launched 1--2\,yr before
core collapse.  However, it is likely that the SN ejecta are slowed by its collision with the CSM, and so it may be more physically relevant to assume
the CSM extent to be $R \gtrsim 9 \times 10^{14}$\,cm (the blackbody radius at $V$-band peak); material at that radius
was launched $<1$\,yr before collapse.

It is not clear whether the extreme mass loss from \thisSN's progenitor was episodic and brief 
or sustained over a year or more, nor is it known whether the assumption of spherical symmetry is appropriate.  
We leave a more thorough examination of these questions to future work.


\subsection{Constraints on Progenitor Variability}

The presence of dense CSM suggests a recent history of extreme mass loss and perhaps variability of the progenitor star.  SN~Ibn 2006jc, for example, underwent a bright outburst 
($M = -14$\,mag) about 2\,yr before becoming a genuine SN \citep{2006CBET..666....1N,2007ApJ...657L.105F,2007Natur.447..829P},
and there are multiple cases where SN~IIn progenitors have been detected in outburst in the years
prior to core collapse \citep[e.g.,][]{2014ApJ...789..104O}, though such outbursts do not appear to be
ubiquitous \citep[e.g.,][]{2015MNRAS.450..246B}.
KAIT has been monitoring NGC~2388 for almost
20\,yr and we searched this extensive dataset for evidence of pre-explosion variability.
We augment our unfiltered KAIT data with the publicly available PTF/iPTF images
of the field in the $r$ passband.\footnote{\url{http://irsa.ipac.caltech.edu/applications/ptf/}}

\begin{figure}
 \includegraphics[width=\columnwidth]{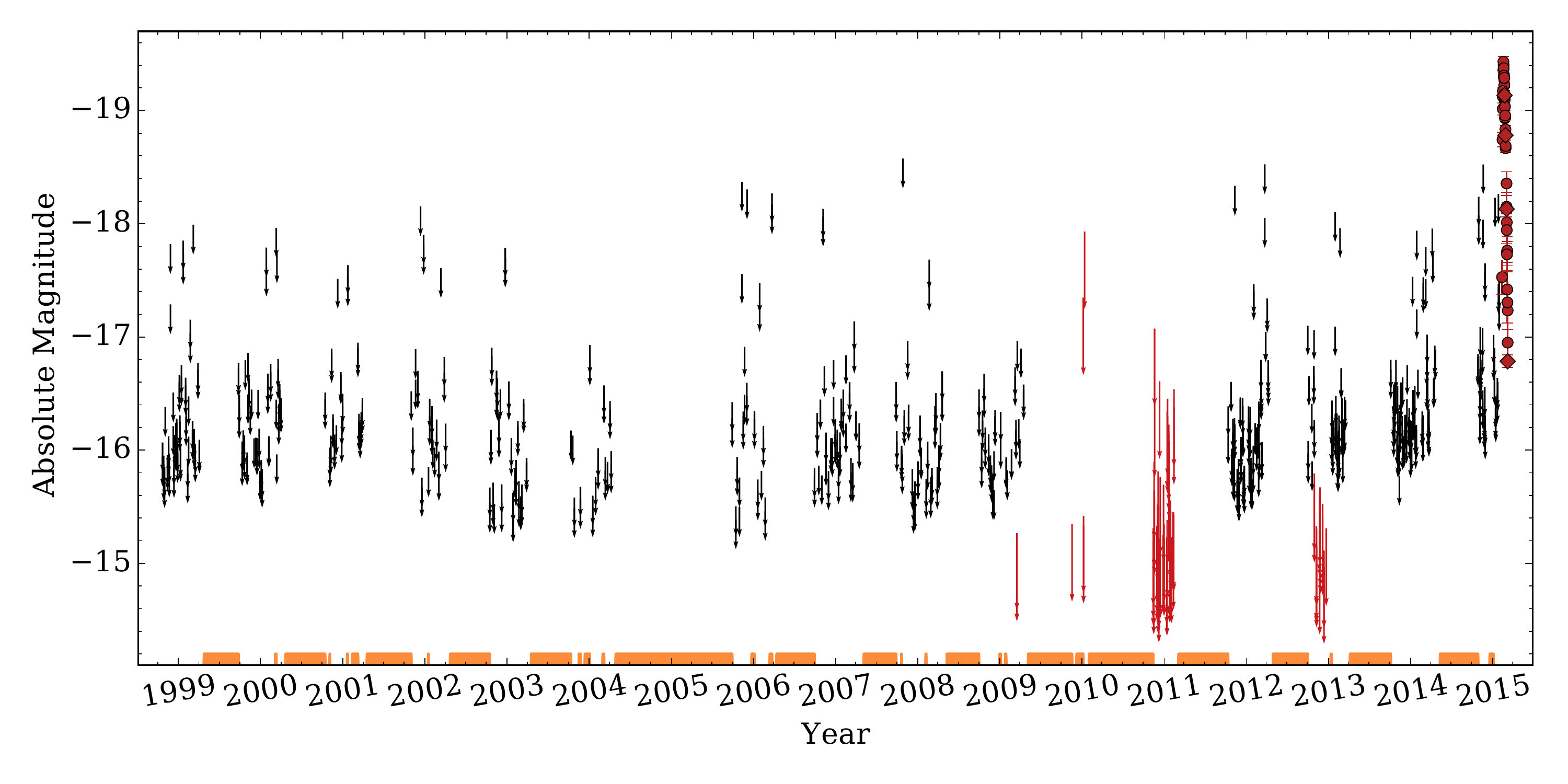}
 \caption{Our 1$\sigma$ nondetections from KAIT unfiltered images (black) and
    PTF/iPTF $R$ images (red).  The light curve of \thisSN\ is shown to the far right.
    Timespans for which no upper limit had been obtained for at least 20 days are
    marked in orange along the bottom.
\label{fig:pre-explosion}}
 \end{figure}

Examining 758 images with observed upper limits fainter than 17.0\,mag from 
1998 Oct. through 2015 Feb. 10, we find no evidence for previous outbursts of the progenitor
based upon difference imaging (using one of our deepest single exposures as a template).
We stacked our images into rolling-window time bins 20\,d wide and subtracted templates to
search for evidence of previous outbursts of that timescale, but found none.  We also stacked all images
together and searched for evidence of a progenitor, but found none.  (Note
that we stacked the KAIT and PTF/iPTF images separately, and we did not perform difference imaging on these very deep stacks,
as we had no templates with which to compare.)
Figure~\ref{fig:pre-explosion} plots our 1$\sigma$ single-image nondetections and the observed light curve of \thisSN.

Though we have regular imaging of NGC~2388 most years since 1998, the field was inaccessible to our
telescopes for several months every year, so there are significant gaps, especially
compared to the timescale of \thisSN's light curve ($\sim20$\,d).  The orange bars along the bottom of
Figure~\ref{fig:pre-explosion} mark every night at which more than 20 days had passed since
the previous upper limit, indicating timespans when a previous \thisSN-like event could have occurred undetected.

Unfortunately, even the epochs that were covered by our monitoring campaigns do not yield stringent constraints on the
luminosity of a previous outburst, partially owing to the estimated $\sim1.5$\,mag of host-galaxy dust extinction along the line of sight toward \thisSN.
The detected outburst of SN~2006jc's precursor reached a peak of $M_r \approx -14.1$\,mag \citep{2007Natur.447..829P}, and
such an outburst in \thisSN's precursor would have probably remained undetected by our monitoring campaign.

\section{Conclusion}
\label{sec:conclusion}

In this paper we presented observations of \thisSN, a highly extinguished, low-velocity, rapidly evolving, luminous, apparently hydrogen-free SN
which exploded in the strongly star-forming galaxy NGC~2388.
Though detailed modeling has yet to be performed, and the degree of host-galaxy dust interference is uncertain,
our data indicate that \thisSN\ was a core-collapse SN with a peak powered by shock breakout
from a dense CSM rather than radioactive decay.  We suggest that this CSM was not wind-like but was
instead created by at least one extreme episode of mass loss
($\dot M \approx 0.2$--1.2\,M$_{\odot}$\,yr$^{-1}$) within a few years of core collapse.
The CSM that surrounded \thisSN\ was effectively hydrogen-free but was helium-rich; we also detect
features from nitrogen, iron, and probably carbon in our spectra.
No long-lasting light-curve tail was observed from radioactivity or from ongoing CSM interaction,
implying that \thisSN\ produced a relatively small amount of $^{56}$Ni compared to normal SNe~Ib/c.

 \thisSN\ is a remarkably well-observed SN~Ibn, especially
among the rapidly fading subset of these events, and we find many similarities between it and other 
SNe in the literature.  Modern surveys indicate the existence of a class of blue, continuum-dominated, hydrogen-deficient, luminous, and rapidly evolving
SNe including those found by \citet{2014ApJ...794...23D} and \citet{2016ApJ...819...35A}; our analysis of \thisSN\ implies fundamental
similarities between these events, a subset of the SNe~Ibn \citep[e.g.,][]{2016MNRAS.456..853P},
and other rapidly fading SNe of lower luminosity (i.e., SN 2002bj and SN 2010X).
While the exact progenitors of these events are quite uncertain, it is clear that they demand extreme mass-loss rates from their 
stripped-envelope progenitor stars.

\section*{Acknowledgements}

The authors thank M.~Drout, R.~Foley, P.~Nugent, D.~Kasen, E.~Quataert, 
D.~Poznanski, and O.~Fox for useful comments and discussions. 
We are grateful to our referee, A.~Pastorello, for comments that have greatly improved this effort.
We thank the {\it HST}, Keck Observatory, and Lick Observatory staffs for their 
expert assistance with the observations.
We also acknowledge the following Nickel 1\,m observers and KAIT checkers
for their valuable assistance with this work:
A.~Bigley, C.~Gould, G.~Halevi, K.~Hayakawa, A.~Hughes,
H.J.~Kim, M.~Kim, P.~Lu, K.~Pina, T.~Ross, S.~Stegman, and H.~Yuk.

A.V.F.'s supernova research group at U. C. Berkeley is supported by NSF
grant AST-1211916, the TABASGO Foundation, Gary and Cynthia Bengier,
and the Christopher R. Redlich Fund. Additional assistance is
provided by NASA/{\it HST} grants AR-14295 and GO-14149 from the
Space Telescope Science Institute, which is operated by the
Association of Universities for Research in Astronomy, Inc., under
NASA contract NAS5-26555. J.M.S. is supported by an NSF Astronomy
and Astrophysics Postdoctoral Fellowship under award AST-1302771.

Some of the data presented herein were obtained at the W. M. Keck
Observatory, which is operated as a scientific partnership among the
California Institute of Technology, the University of California, and
NASA; the observatory was made possible by the generous financial
support of the W. M. Keck Foundation.

KAIT and its ongoing operation
were supported by donations from Sun Microsystems, Inc., the
Hewlett-Packard Company, AutoScope Corporation, Lick Observatory, the
NSF, the University of California, the Sylvia and Jim Katzman
Foundation, and the TABASGO Foundation.  Research at Lick Observatory
is partially supported by a very generous gift from Google, as well as by
contributions from numerous individuals including 
Eliza Brown and Hal Candee, 
Kathy Burck and Gilbert Montoya, 
David and Linda Cornfield, 
William and Phyllis Draper,
Luke Ellis and Laura Sawczuk, 
Alan and Gladys Hoefer,
DuBose and Nancy Montgomery, 
Jeanne and Sanford Robertson,
Stanley and Miriam Schiffman, 
Thomas and Alison Schneider, 
the Hugh Stuart Center Charitable Trust,
Mary-Lou Smulders and Nicholas Hodson,
Clark and Sharon Winslow, 
Weldon and Ruth Wood,
and many others. 

This research was
based in part on data taken with the NASA/ESA {\it Hubble Space Telescope}, and
obtained from the Hubble Legacy Archive, which is a collaboration
between the Space Telescope Science Institute (STScI/NASA), the Space
Telescope European Coordinating Facility (ST-ECF/ESA), and the
Canadian Astronomy Data Centre (CADC/NRC/CSA).  We also made use 
of the NASA/IPAC Extragalactic Database (NED) which is operated by 
the Jet Propulsion Laboratory, California Institute of Technology, 
under contract with NASA.


\bibliographystyle{mnras}
\bibliography{bib}



%
%


\bsp	
\label{lastpage}
\end{document}